\documentclass[fleqn,usenatbib]{mnras}

\usepackage{newtxtext,newtxmath}

\usepackage[T1]{fontenc}

\DeclareRobustCommand{\VAN}[3]{#2}
\let\VANthebibliography\thebibliography
\def\thebibliography{\DeclareRobustCommand{\VAN}[3]{##3}\VANthebibliography}

\usepackage{graphicx}	
\usepackage{amsmath}	
\usepackage{booktabs}
\usepackage{threeparttable}
\usepackage{orcidlink}
\usepackage{soul}

\newcommand{\revision}[1]{\textcolor{black}{#1}}

\title[He absorption in exoplanet atmospheres]{Understanding what helium absorption tells us about atmospheric escape from exoplanets}

\author[Ballabio \& Owen]{
Giulia Ballabio$^{\orcidlink{0000-0002-4687-2133}1}$\thanks{E-mail: g.ballabio@imperial.ac.uk}
and James E. Owen$^{\orcidlink{0000-0002-4856-7837}1,2}$
\\
$^{1}$Imperial Astrophysics, Imperial College London, Blackett Laboratory, Prince Consort Road, London SW7 2AZ, UK\\
$^{2}$Department of Earth, Planetary, and Space Sciences, University of California, Los Angeles, CA 90095, USA
}

\date{Accepted 2025 January 5. Received 2024 December 19; in original form 2024 October 8}

\pubyear{2024}

\begin{document}
\label{firstpage}
\pagerange{\pageref{firstpage}--\pageref{lastpage}}
\maketitle

\begin{abstract}
Atmospheric escape is now considered the major contributing factor in shaping the demographic of detected exoplanets. However, inferences about the exoplanet populations strongly depend on the accuracy of the models. Direct observational tests of atmospheric models are still in their infancy. Helium escape from planetary atmospheres has rapidly become the primary observational probe, already observed in $\gtrsim$20 exoplanets. Grounding our understanding in the basic physics of atmospheric escape, we present a new theoretical model to predict the excess absorption from the helium absorption line. We constrain the atmosphere properties, such as mass-loss rates and outflow temperatures, by implementing a Parker wind solution with an energy limited evaporating outflow. Importantly, we self-consistently link the mass-loss rates and outflow temperatures, which are critical to understanding helium absorption as the triplet-level population is typically exponentially sensitive to temperature. Furthermore, helium absorption is typically optically thin and the absorption is dominated far from the planet. Therefore, the absorption depth is not a measure of the size of the helium outflow. 
\revision{Our results indicate that for planets with a detectable signal, typically the helium triplet population in the atmosphere rapidly approaches a statistical equilibrium between populations by recombination and depopulation caused by electron collisions.}
We suggest that excess helium absorption can be quantified by a scaled equivalent width, which is positively correlated with the mass loss rate. We also show that the helium absorption scales with incident radiation, particularly with the XEUV to FUV flux ratios. 
\end{abstract} 

\begin{keywords}
planets and satellites: atmospheres -- planets and satellites: physical evolution -- radiative transfer
\end{keywords}



\section{Introduction}\label{sec:intro}

Atmospheric escape is believed to significantly influence the distribution of observed exoplanets \citep[e.g.][]{Owen2019}. It is often invoked to explain the absence of short-period Neptunes \citep{Lundkvist2016,2016A&A...589A..75M,OwenLai2018} and the bimodal feature in the radius distribution of Sub-Neptunes and Super-Earths, know as the ``Radius Valley'' \citep{2017AJ....154..109F, Owen17, 2018AJ....156..264F, 2018MNRAS.479.4786V}. 

The first proposed mechanism for atmospheric escape driven evolution is photoevaporation \citep[e.g.][]{Baraffe2005,Lopez2012,2013ApJ...775..105O, 2013ApJ...776....2L, 2016ApJ...831..180C,2021MNRAS.503.1526R,Rogers2023}. High-energy ultraviolet (UV) radiation from the star ionizes the atoms in the planet's upper atmosphere, which can then acquire enough thermal energy to escape the planet's gravitational pull \citep[e.g.][]{Lammer2003,MC2009}. For close-in planets, the entire upper atmosphere can often be heated by the absorption of stellar X-rays and extreme UV radiation to such an extent that it creates a strong outflow of gas \citep[e.g.][]{Yelle2004,GarciaMunoz2007,Owen_Jackson_12}. This outflow is then strong enough that it can impact a planet's atmosphere over evolutionary timescales \citep[e.g.][]{Lecavelier2007,Jackson2012,Owen2019}.

Recently, an alternative mechanism for atmospheric escape has been proposed: ``core-powered mass-loss'', wherein the planet's cooling core and the star's bolometric radiation heat the atmosphere, facilitating its escape \citep{2018MNRAS.476..759G, 2019MNRAS.487...24G, 2022MNRAS.516.4585G}. Similar to the photoevaporative case, this heating drives an outflow, resulting in substantial mass-loss over evolutionary timescales \citep[e.g.][]{Gupta2020,Misener2024}. While both photoevaporation and core-powered mass-loss typically produce a transonic wind, they differ notably in the final properties of the outflow, such as the mass-loss rates and the wind temperatures, likely producing different observable signatures \citep[e.g.][]{Gupta2021,Owen2023,SchreyerLy2024}. 

While recent theoretical work has suggested that photoevaporation and core-powered mass-loss play a role in shaping the exoplanet population \citep{Owen2024,Rogers2024}, atmospheric escape models are yet to be directly, quantitatively and systematically tested observationally. A major reason for this slow progress is not a limited number of observations but a lack of physical understanding of what observations of atmospheric escape are actually measuring \citep[e.g.][]{dosSantos2023}. Recently, the first observational tracer of atmospheric escape \citep[e.g.][]{2003Natur.422..143V}, Ly-$\alpha$ transits, has been put on a more firm theoretical foundation. \citet{SchreyerLy2024} showed that the outflow's sound speed could be extracted from observations, potentially distinguishing between core-powered mass-loss and photoevaporation; however, it was only weakly sensitive to the mass-loss rate. Furthermore, Ly-$\alpha$ transits only probe neutral hydrogen and thus becomes transparent at high fluxes as the outflow is ionized \citep[e.g.][]{Owen2023}. Given the mass-loss rate is highest at high fluxes, this means it is difficult to use Ly-$\alpha$ transits to probe the most important phase of atmospheric escape-driven evolution, as evidenced by recent non-detections for highly irradiated young planets \citep[e.g.][]{Rockcliffe2021,Zhang2022,Morrissey2024,Alam2024}.

A significant advancement in observing ongoing mass-loss was achieved with the first detection of helium in an exoplanet’s extended atmosphere \citep{2018Natur.557...68S}, indicating the atmosphere was escaping. Helium, specifically the triplet absorption feature at 10830\AA~ in the near-infrared, offers substantial observational advantages over previous UV methods \citep[e.g.][]{Seager2000,2018ApJ...855L..11O}. 
Given the 10833\AA~ triplet's accessibility from the ground, at high spectral resolution \citep[e.g.][]{2018Sci...362.1384A}, the study of helium escape from exoplanet atmospheres, therefore, provides an important new window into atmospheric dynamics and planetary evolution. 

Since 2018, a rapidly growing number of studies have focused on exoplanetary helium \citep[e.g., ][]{2018Sci...362.1388N, 2020AJ....159..115K, 2020AJ....159..278V, Spake2021, 2022AJ....164...24K, 2023AJ....165...62Z, 2023ApJ...953L..25Z}, leading to more than ten confirmed detections of helium in various exoplanets \citep[e.g.][]{dosSantos2023,2024arXiv240416732O}. Thus, helium has rapidly become the dominant observational tracer of an exoplanet's escaping atmosphere.  The sample of detected planets reveals that K-type and M-type stars are particularly favourable for observing helium. This result is consistent with the fact the helium triplet metastable state is efficiently populated at high extreme-UV (EUV) to mid-UV flux ratios \citep{2019ApJ...881..133O}. Previous studies have shown that exoplanets subjected to high X-ray and UV irradiation exhibit more pronounced helium absorption features \citep[e.g., ][]{2018Sci...362.1388N, 2019A&A...629A.110A}, and are likely sensitive to planetary magnetic fields \citep[e.g.][]{Oklopcic2020,SchreyerHe2024}. Subsequently, \cite{2022MNRAS.512.1751P} reviewed a selection of studies on He~I absorption and found that the strength of the helium absorption correlates more closely with the EUV fluxes. 

Current modelling efforts have used simulations to predict the helium transit depth. \citet{2018ApJ...855L..11O} used an isothermal Parker wind, along with a minimal model of the helium level populations, to predict the transit depth, successfully explaining that larger transit depths should be observed around K-type hosts \citep{2019ApJ...881..133O}. This idea has been developed further into the {\sc pwinds} code \citep{dosSantos2022} that is commonly used to fit observations \citep[e.g.][]{Alam2024}. Recent simulation efforts have focused on expanding the model for the helium level populations \citep[e.g.][]{2024MNRAS.527.4657A} and highlighting the critical role the temperature plays in setting the helium signal \citep[e.g.][]{Linssen2022,biassoni2024} leading to more advanced fitting tools \citep[e.g.][]{Linssen2024}. Furthermore, tracking the helium level populations has been incorporated into multi-dimensional radiation-hydrodynamic simulations, allowing more detailed predictions of the transit profiles \citep[e.g.][]{MacLeod2022,SchreyerHe2024}. 

However, while these simulation works have allowed an understanding of how various planetary and stellar parameters correlate with the strength of the helium signal, along with detailed predictions of the line profiles, they do not provide a physical picture of what helium 10830 \AA~ transits actually constrain. Guided by previous simulation results, in this work, we aim to understand from a first principled approach what helium transits actually measure, uncovering their sensitivity to the multitude of planetary and stellar parameters. We do this by studying a minimal analytic model in Section~\ref{sec:overview}, which elucidates the basics, followed by a complete semi-analytic model in Section~\ref{sec:better_model}.

\section{Overview of Helium Escape} \label{sec:overview}

Before we build a more physically complete model, it's worth considering the absolute basics of how the helium 10830~\AA~ triplet behaves to provide a physical grounding on which to base our further discussion. It is worth emphasising at this point that given helium is observed in transmission, then in an escaping, extended atmosphere, the approximation that the planet's atmospheric scale height ($H$) is considerably smaller than the planet's radius ($R_p$) no longer holds. Thus, much of the intuition that arises from typical transmission spectroscopy, where the limb optical path is primarily sensitive to a few scale heights around the slant photosphere \citep[e.g.][]{Seager2000,Fortney2005}, no longer holds. In fact, helium transmission spectroscopy is most sensitive to the diffuse, optically thin outer regions of the outflow. This result was already seen in \citet{2018ApJ...855L..11O}, where different choices of the simulated outer boundary resulted in different transit depths.

When considering the transmission spectra of a transiting exoplanet in the helium 10830\AA~ line, the optical depth along the line of sight $z$ with impact parameter $b$ is calculated as,
\begin{equation}
    \tau(b) = 2 \int_0^\infty \sigma_0 \, n_3(r) dz, 
    \label{eq:optical_depth_abs}
\end{equation}
where $n_3(r)=f_3(r) n_{\rm He}(r)$, with $f_3(r)$ being the fraction of helium in the triplet state and $n_{\rm He}(r)=f_{\rm He} n(r)$ is the radial number density profile of helium, given a helium number fraction, $f_{\rm He}$ and total number density, $n$.\footnote{Fractionation can result in $f_{\rm He}$ being spatially variable; however, we neglect this complication here.} The absorption cross-section $\sigma_0$ is given by
\begin{equation}
    \sigma_0 = \frac{\pi e^2}{m_e c} \zeta, 
\end{equation}
where $e$ and $m_e$ are the electron's charge and mass, respectively. $c$ is the speed of light and $\zeta$ the oscillator strength of the helium transition at 10830~\AA. Now, rewriting Eq.~\ref{eq:optical_depth_abs} as a function of radius from the planet's centre, $r$, we obtain
\begin{equation}
    \tau(b) = 2 \int_b^\infty \sigma_0 \, f_3(r) f_{\rm He} \, n(r) \frac{r dr}{\sqrt{r^2 - b^2}}. 
\end{equation}
As we have already mentioned, and will demonstrate \revision{in Appendix~\ref{appx:optical_depth_absorption}}, helium absorption is dominated at large radii from the planet. Thermally launched winds like those that occur in atmospheric escape have a slowly (logarthimically) increasing velocity with distance at large radii \citep[e.g.][]{1958ApJ...128..664P,Lamers1999}. A result that has been borne out by numerical simulations of exoplanet mass-loss in both photoevaporation \citep[e.g.][]{Yelle2004,MC2009,Schulik2023} and core-powered mass-loss \citep[e.g.][]{Misener2024}. 
Thus, assuming for now, that the velocity in the outflow is constant, for constant mass-loss rate, the density $n(r)$ drops as $r^{-2}$, yielding
\begin{equation}
    \tau (b) = 2 \int_b^\infty \sigma_0 \, f_3(r) \, f_{\rm He} \, n_0 \, R_{\rm p}^2 \, \frac{dr}{r \sqrt{r^2 - b^2}},
    \label{eq:tau_r}
\end{equation}
where $n_0$ is the density at the base of the outflow (taken to be the planet's radius $R_{\rm p}$). The outflow is typically sufficiently dense so that recombinations into the triplet state are balanced by collisional excitations from the triplet to the singlet state (at least for K and M stars, which typically dominate the observations), and the fraction of helium in the triplet can be simplified as \citep{2019ApJ...881..133O,biassoni2024}:
\begin{equation}
    f_3(r) \approx \frac{\alpha_3(T)}{q_{3s}(T) + q_{3p}(T)} \left[1-f_1(r)\right], 
    \label{eq:f_he_3}
\end{equation}
where $\alpha_3$ is the recombination rate to the triplet state and $q_{3s}$ and $q_{3p}$ are the collision rates from the triplet to singlet state and are all functions of the wind temperature (see discussion in Section~\ref{sec:better_model} for more details). 
Since $f_3 \ll 1$, then in advection/ionization equilibrium, the fraction of helium in the singlet state can be derived by solving \citep{2018ApJ...855L..11O}:
\begin{equation}
    \varv \frac{\partial f_1}{\partial r} = - f_1 \Phi_1 e^{-\tau_1} , 
\end{equation}
where $\Phi_1$ is the helium photoionization rate and $\tau_1$ is the optical depth (calculated using the flux-averaged cross section).
Since we are concerned with absorption at large radii, the gas is optically thin to helium ionizing photons, and using our constant velocity assumption, we find: 
\begin{equation}
    f_1(r) = {\rm exp}\left(-\int_{R_{\rm p}}^{r} \frac{\Phi_1}{\varv}  e^{-\tau_1} dr \right) \approx {\rm exp}\left[-\frac{\Phi_1 }{\varv} (r - R_{\rm p}) \right] ,
    \label{eq:f_he_1}
\end{equation}
Combining Eq.~\ref{eq:tau_r}, \ref{eq:f_he_3} and \ref{eq:f_he_1}, we obtain
\begin{multline}
    \tau(b) \approx 2 \, \sigma_0 f_{\rm He} \, n_0 \, R_{\rm p}^2 \, {\rm K}(T) \\
    \int_b^\infty \left\{1 - {\rm exp}\left[-\frac{\Phi_1 }{\varv} (r - R_{\rm p}) \right] \right \} \frac{dr}{r \sqrt{r^2 - b^2}}, 
\end{multline}
with the function ${\rm K}(T)$ defined as:
\begin{equation}
    {\rm K}(T) = \frac{\alpha_3(T)}{q_{3s(T)} + q_{3p(T)}}.
\end{equation}
capturing the atomic physics and highlighting that it is only a function of temperature. In the limit in which the exponential term is small, and adopting an isothermal outflow, the integral can be easily solved, yielding
\begin{equation}
    \tau(b) \approx \sigma_0 f_{\rm He} \, n_0 \, R_{\rm p}^2 \, {\rm K}(T) \frac{\pi}{b}.
    \label{eq:final_tau}
\end{equation}
\revision{This approximation is verified using our numerical models in Appendix~\ref{appx:main_reactions}.}
\begin{figure}
    \centering
    \includegraphics[width=0.85\columnwidth,trim=0cm 0cm 0cm 0cm,clip]{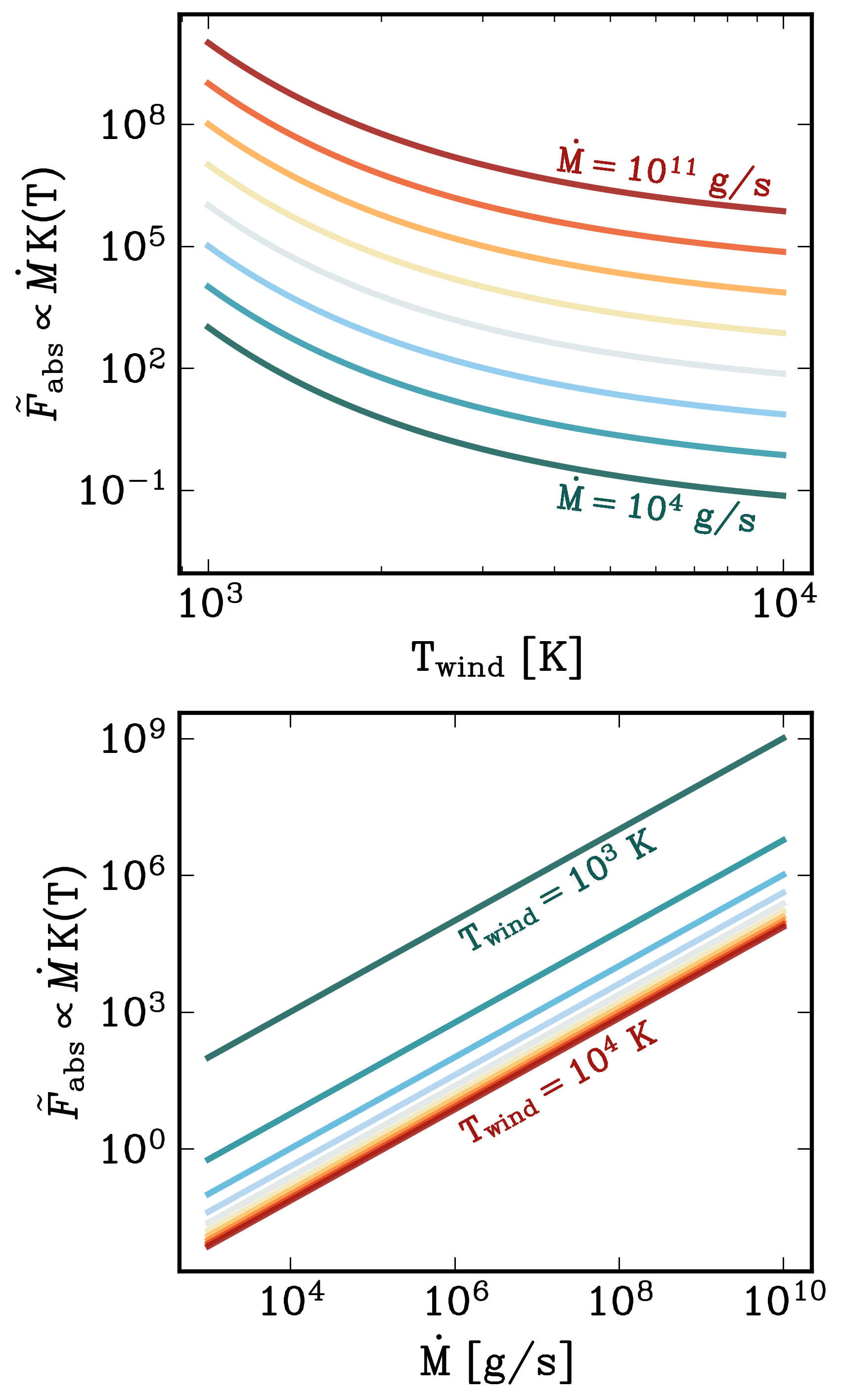}
    \caption{\revision{Theoretical expectation of the scaled excess absorption as a function of both mass-loss rate and wind temperature. Each curve represents the line of constant mass-loss rate (top panel) or constant temperature (bottom panel). The scaled equivalent width has arbitrary units, as only the dependency on those parameters is shown here.}}
    \label{fig:theoretical_behaviour}
\end{figure}
Helium transits are often characterised in terms of ``excess absorption'' \citep[e.g.][]{2020AJ....159..115K}, which is the absorption measured in the 10830~\AA~line in excess of the planet's broadband absorption outside the line. Thus, the excess absorption along the line of sight is calculated as
\begin{equation}
    F_{\rm abs} = \int^{R_{\rm out}}_{R_{\rm p}} \frac{2 \pi b}{\pi R^2_*} \left[1 - e^{-\tau(b)}\right] db, 
    \label{eq:excess_absorption}
\end{equation}
where $R_{\rm out}$ is the outer extent of the outflow. The excess absorption is normalized to the "total flux" of the star, $\pi R^2_*$, where $R_*$ is the stellar radius. 
Thus, in the case of optically thin absorption and combining Eq.\ref{eq:final_tau}, we find
\begin{equation}
    F_{\rm abs} \approx 2 \pi \sigma_0 f_{\rm He} \, n_0 \, {\rm K}(T) \, R_{\rm p}^2 \frac{ \left(R_{\rm out} - R_{\rm p} \right)}{R_*^2} .
\end{equation}
It is worth pausing here to note that the excess absorption does not converge with increasing $R_{\rm out}$. In fact, any density profile that falls off shallower than $1/r^3$ will not converge. Thus, transmission spectroscopy of extended atmospheres generally occurs as optically thin absorption rather than probing the impact parameter of the optically thick/thin transition in classical transmission spectroscopy. This has two important consequences: i) converting a measured transit depth into the obscuring ``radius'' does not provide a representative size of the outflow, and it is, in fact, significantly larger than this radius; ii) the choice of the size of the outflow is important in setting the absorption. Obviously, $R_{\rm out}$ cannot be larger than the size of the star; however, this is not generally the main determinant of the outflow size. A planetary outflow will expand quasi-spherically until tidal forces (or interactions with the stellar wind \citealt{McCann2019,MacLeod2022}) confine the outflow. Thus, the maximum extent of the outflow is when the Coriolis force bends the streamlines back on themselves, limiting the outflow size to the Coriolis turning length \citep{McCann2019,Owen2023}:
\begin{equation}
    R_{\rm c} = \frac{c_s}{2\Omega}
\end{equation}
where $c_s$ is the isothermal sound speed of the outflow, and $\Omega$ is the planet's angular velocity. \revision{We want to underline that this is an approximation of the model, the outflow could be compressed by the stellar wind and there are a handful of cases for which helium absorption extends beyond the Coriolis radius, forming leading or trailing tails \citep[e.g.][]{MacLeod2022}. Some of the examples present highly asymmetric helium light curves, either showing strong leading arms \citep[e.g. HAT-P-32b or HAT-P-67b;][]{2023SciA....9F8736Z, 2024AJ....167..142G} or trailing tails \citep[e.g. WASP-107b or WASP-69b;][]{2021AJ....162..284S, 2024ApJ...960..123T}.}

Using the Coriolis length as our maximum radius, and following mass conservation $\dot{M} = 4\pi n_0 m_{\rm H} c_s R_{\rm p}^2$, 
\begin{equation}
    F_{\rm abs} = \sigma_0 \, {\rm K}(T) \frac{f_{\rm He}}{2 m_{\rm H}} \frac{\dot{M}}{c_s} \, \frac{ \left(R_{\rm c} - R_{\rm p} \right)}{R_*^2} ,
\end{equation}
where $m_{\rm H}$ is the mass of the hydrogen atom. 
In the limit of $R_{\rm c}\gg R_{\rm p}$, as typical, we can simplify the expression even further
\begin{equation}
    F_{\rm abs} = \sigma_0 \, {\rm K}(T) \frac{f_{\rm He}}{2 m_{\rm H}} \frac{\dot{M}}{2 \Omega R_*^2} ,
    \label{eq:final_excess_abs}
\end{equation}
We now rescale the excess absorption by the ``\textit{geometric} factor'', $1/(\Omega R_*^2)$, which has the inverse units of a specific angular momentum~\footnote{As a consequence, the scaled excess absorption has the units of a specific angular momentum, so we normalise it to the Earth's specific angular momentum $\ell_{\oplus}$ for simplicity.},
and obtain the scaled excess absorption:
\begin{equation}
    \widetilde{F}_{\rm abs} = \dot{M} \, \frac{f_{\rm He} \sigma_0}{4 m_{\rm H}} {\rm K}(T).
    \label{eq:scaled_excess_abs}
\end{equation}
Thus, we have demonstrated from first principles that the helium triplet excess absorption scales linearly with the atmospheric mass-loss rate, with all other things being equal. Unlike Ly-$\alpha$,  helium absorption detection can, therefore, be used as a direct measure of the mass-loss rate from an exoplanetary atmosphere. We plot the expected equivalent width as described in Eq.~\ref{eq:scaled_excess_abs}, in Figure~\ref{fig:theoretical_behaviour}. The top panel shows that for a constant mass-loss rate, the equivalent width decreases with increasing temperature. 
This is because collisional de-excitations, which depopulate the helium triplet state, dominate over recombinations with increasing temperature. \revision{Higher temperatures in the outflow also lead to a faster wind and a lower gas density, which contributes to produce a lower absorption signal.}
Inversely, lines of constant temperature show an increasing equivalent width with an increasing mass-loss rate, as illustrated in the bottom panel. Intuitively, a higher mass-loss rate will result in a deeper transit signal from helium absorption. However, as pointed out by \citet{biassoni2024}, the exponential dependence of the triplet's collisional depopulation rates results in an exponential sensitivity of the helium signal on temperature. Therefore, despite having a direct and linear dependence on mass-loss rate, the outflow temperature will play an important role in setting the observed transit depth. 

Despite being illustrative, we have made a few assumptions in our derivation to simplify our calculations. In particular, we have not physically linked the mass-loss rate and outflow temperature. In the next section, we now construct an atmospheric model to where we physically link the outflow temperature and mass-loss rate, given the importance both play in setting the helium absorption signal.

\section{Atmospheric Escape Modelling} \label{sec:better_model}
We consider a hydrogen and helium-dominated exoplanet atmosphere and construct a 1D model for the escaping gas, and numerically compute the excess absorption in the He absorption line for a population of Sub-Neptunes, Neptunes and Jupiters-like planets. The goal here is to move beyond the assumption that the temperature and mass-loss rate are independent, as typically assumed in previous modelling efforts. Rather, we will use the physics of atmospheric escape to directly link the temperature and mass-loss rate.

\subsection{Atmosphere Structure} \label{sec:atmosphere_structure}
\begin{figure}
    \centering
    \includegraphics[width=0.85\columnwidth,trim=0cm 0cm 0cm 0cm,clip]{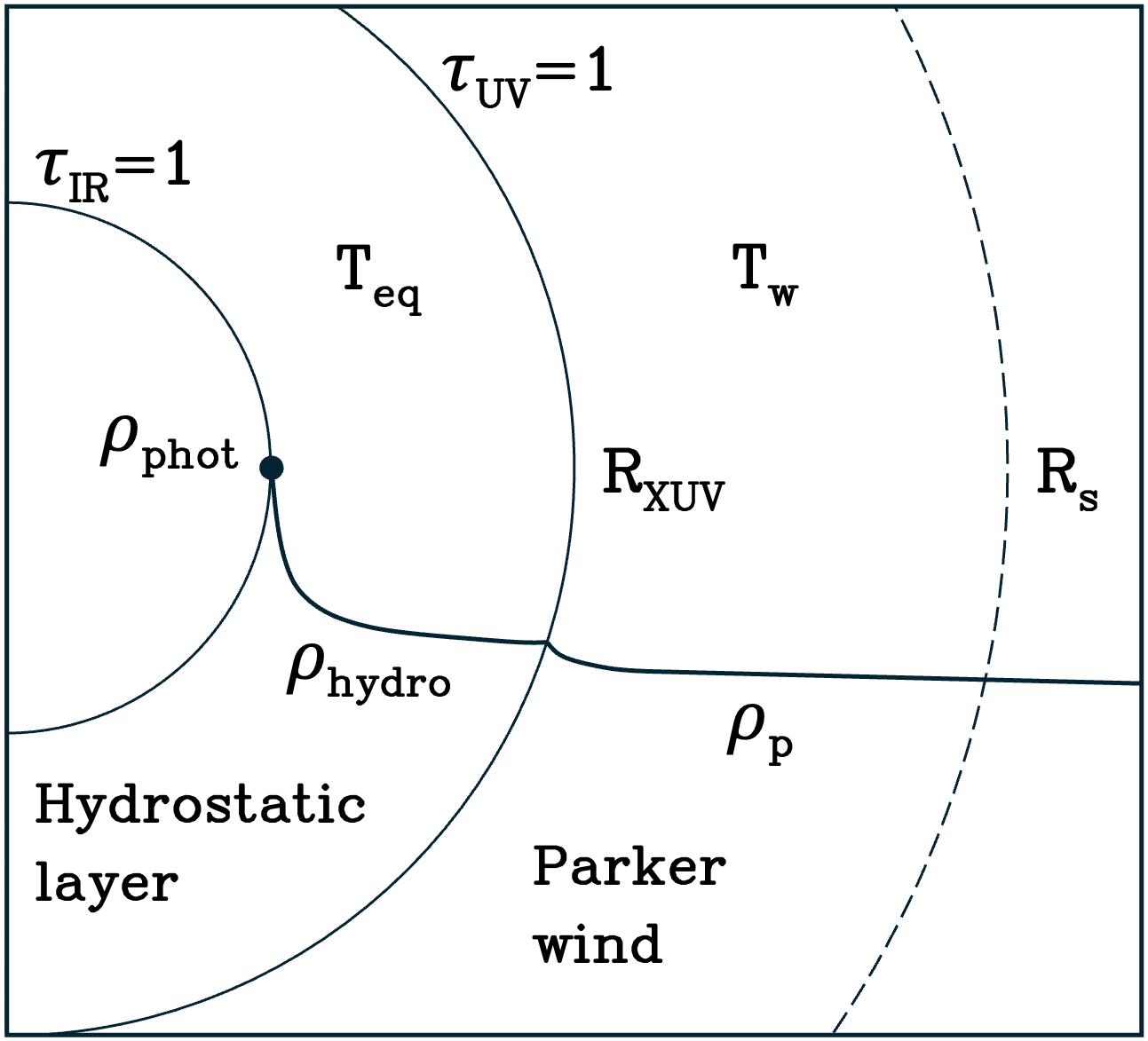}
    \caption{Our 1D model for the planet's atmosphere. We model the atmosphere as a two-layer structure: the bottom layer in hydrostatic equilibrium and the top layer as an isothermal Parker wind. The solid line indicates the radial density profile in the two layers. The distances are not scaled.}
    \label{fig:atmosphere}
\end{figure}

Following \citet{Owen2024}, we model the planet's atmosphere and outflow as a two-layer structure. The base layer is composed of cool gas in hydrostatic equilibrium, while the uppermost layer contains hot gas and is modelled as an isothermal Parker wind \citep{1958ApJ...128..664P, Lamers1999}, as illustrated by the sketch in Figure~\ref{fig:atmosphere}. 

For a time-independent and spherically symmetric wind, the mass conservation equation yields
\begin{equation}
    \Dot{M} = 4 \pi r^2 \rho(r) \varv(r), 
    \label{eq:mass_cons}
\end{equation}
where $r$ is the height of the atmosphere, $\rho(r)$ and $\varv(r)$ are the gas density and velocity, respectively. The momentum conservation equation is
\begin{equation}
    \varv(r) \frac{d\varv}{dr} + \frac{1}{\rho(r)} \frac{dP}{dr} + \frac{G M_{\rm p}}{r^2} - \frac{3 G M_* r}{a^3}= 0,
    \label{eq:mom_cons}
\end{equation}
where $P$ is the gas pressure and $M_{\rm p}$ is the planet's mass. $M_*$ is the stellar mass and $a$ is the orbital separation. 
The acceleration of the gas is driven by the balance between the pressure gradient and the gravitational force, including the tidal effects. We now solve Eq.~\ref{eq:mom_cons} for the two layers. 

In hydrostatic equilibrium, $d\varv/dr=0$, and the solution to Eq.~\ref{eq:mom_cons}, gives a radial density profile of the form: 
\begin{equation}
    \frac{\rho_{\rm hs}(r)}{\rho_{\rm phot}} = {\rm exp}\left\{ \frac{G M_{\rm p}}{c_{s, {\rm eq}}^2} \left( \frac{1}{r} - \frac{1}{R_{\rm p}} \right) \right\}, 
    \label{eq:hydrostatic_density}
\end{equation}
 where being well inside the Hill sphere of the planet, we can neglect the tidal effects, as the planet's gravity dominates the star's gravity for this cool gas. The sound speed $c_{s, {\rm eq}}$ is defined as
\begin{equation}
    c_{s, {\rm eq}} = \sqrt{\frac{\kappa_B T_{\rm eq}}{\mu m_{\rm H}}},
    \label{eq:sound_speed_eq}
\end{equation}
where we set the temperature of this cool gas to  $T_{\rm eq}$, the planet's equilibrium temperature. \revision{A gas mixture dominated by molecular hydrogen has a mean molecular weight $\mu \approx 2.35$, which is assumed constant throughout the bolometric region.}
The density $\rho_{\rm phot}$ is estimated at the planet's photosphere, where the gas transitions from optically thick to optically thin to the outgoing infrared radiation.
\begin{equation}
    \rho_{\rm phot} \approx \frac{g}{c_{s, {\rm eq}}^2 \kappa_{\rm IR}} 
\end{equation}
where $g$ is the gravitational acceleration at the planet's photosphere and $\kappa_{\rm IR} = 10^{-2} ~{\rm cm}^2 {\rm g}^{-1}$ is the opacity to the IR radiation, an appropriate value for close-in hydrogen/helium dominated atmospheres \citep[e.g.][]{Guillot2010,Freedman2014}.

\begin{figure}
    \centering
    \includegraphics[width=0.95\columnwidth,trim=0cm 0cm 0cm 0cm,clip]{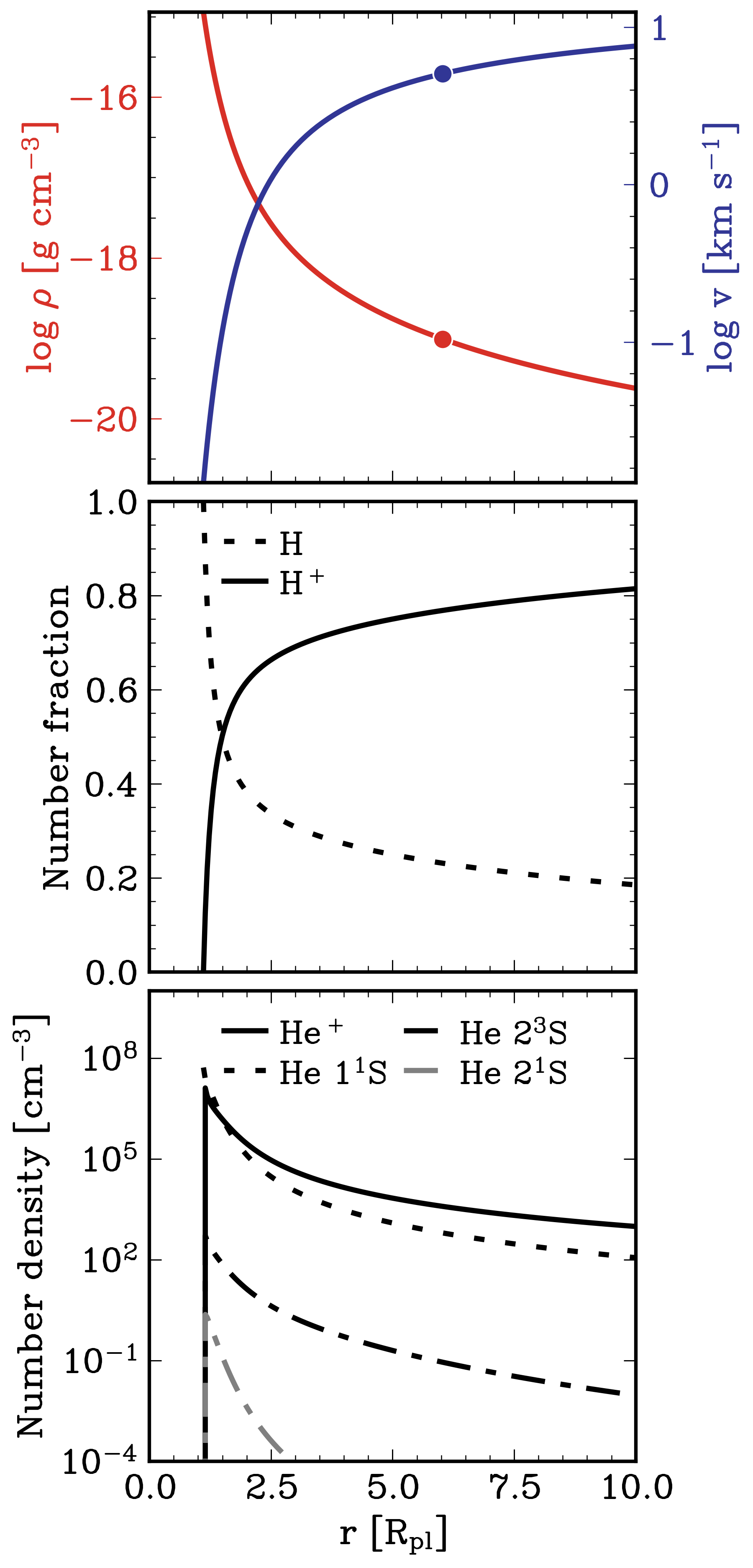}
     \caption{\textit{Top:} Typical 1D Parker wind structure shown for a GJ 436b analog, as an example, \revision{orbiting an M2.5 star at a distance of 0.05 au. The model gives an outflow temperature of 3350~K and a mass-loss rate of $1.72 \times 10^8$~g~s$^{-1}$.} 
     The radial velocity (blue) and density (red) profiles are derived from Eq.~\ref{eq:vel_parker} and \ref{eq:dens_parker}, respectively. 
     The dots indicate the location of the sonic point. \textit{Middle:} Radial number fractions of neutral and ionized hydrogen, derived from Eq.\ref{eq:dfion_dr}. \textit{Bottom:} Radial number density profiles of the helium states, derived from Eqs.~\ref{eq:df1_dr}, \ref{eq:df3_dr} and \ref{eq:dfs_dr}.}
    \label{fig:profiles}
\end{figure}


Assuming an isothermal outflow, the energy equation reduces to the simple form $T(r) = {\rm constant}$, which implies
\begin{equation}
    P = \rho c_s^2,
    \label{eq:energy_cons}
\end{equation}
where the sound speed is
\begin{equation}
    c_s = \sqrt{\frac{\kappa_B T_{\rm wind}}{\mu m_{\rm H}}}.
    \label{eq:sound_speed_w}
\end{equation}
This refers to the sound speed of the gas in the outflow region, where $T_{\rm wind}$ is the outflow temperature. \revision{In the outflow region, the mean molecular weight is $\mu = 1.08$, assuming a gas dominated by neutral atomic hydrogen, with a helium fraction of 10\%.}
In the Parker wind model, the gas is slow close to the planet's surface and gradually accelerates due to the pressure force until it reaches the sonic radius and becomes supersonic. The location of the sonic point is given by:
\begin{equation}
    R_s = \frac{G M_{\rm p}}{2 c_s^2}.
    \label{eq:sonic_radius}
\end{equation}
We note here that in principle, we can define two sonic radii: a \textit{hot} sonic radius in the outflow region and a \textit{cold} sonic radius in the hydrostatic layer, using Eq.~\ref{eq:sound_speed_eq}. However, the gas in hydrostatic equilibrium would only become supersonic in the case of core-powered mass-loss \citep{Owen2024}. As a helium signal requires ionizing photons (since it's formed by recombination), then a core-powered mass-loss outflow does not yield a detectable helium signal \citep{2023AJ....165...62Z}, a result we will explicitly demonstrate. Thus, our sonic radius refers specifically to the \textit{hot} sonic radius. 
Following \cite{Lamers1999}, we can find a solution to Eq.~\ref{eq:mom_cons} for the radially expanding atmosphere. The velocity profile is given by
\begin{equation}
    \frac{\varv_{\rm p}(r)}{c_s} {\rm exp} \left\{ -\frac{\varv_{\rm p}(r)^2}{2 c_s^2} \right\} = \left( \frac{R_s}{r}\right)^2 {\rm exp} \left\{ -\frac{2 R_s}{r} + \frac32 \right\}.
    \label{eq:vel_parker}
\end{equation}
Combining Eq.~\ref{eq:vel_parker} with Eq.~\ref{eq:mass_cons}, we obtain the density profile 
\begin{equation}
    \frac{\rho_{\rm p}(r)}{\rho_s} = {\rm exp} \left\{ \frac{2 R_s}{r} - \frac32 - \frac{\varv_{\rm p}(r)^2}{2 c_s^2} \right\},
    \label{eq:dens_parker}
\end{equation}
where $\rho_s$ is the density at the sonic point. The solutions to the Parker wind model for GJ 436b (as an illustrative example) are shown in the top panel of Figure~\ref{fig:profiles}.

The problem has a few free parameters: known parameters such as the planet's mass and radius and the planet's equilibrium temperature; whereas some currently, as yet, unconstrained parameters such as the wind temperature and the mass-loss rate. Thus, we make some additional considerations to reduce the number of free parameters. 

\begin{figure*}
    \centering
    \includegraphics[width=0.7\textwidth,trim=0cm 0cm 0cm 0cm,clip]{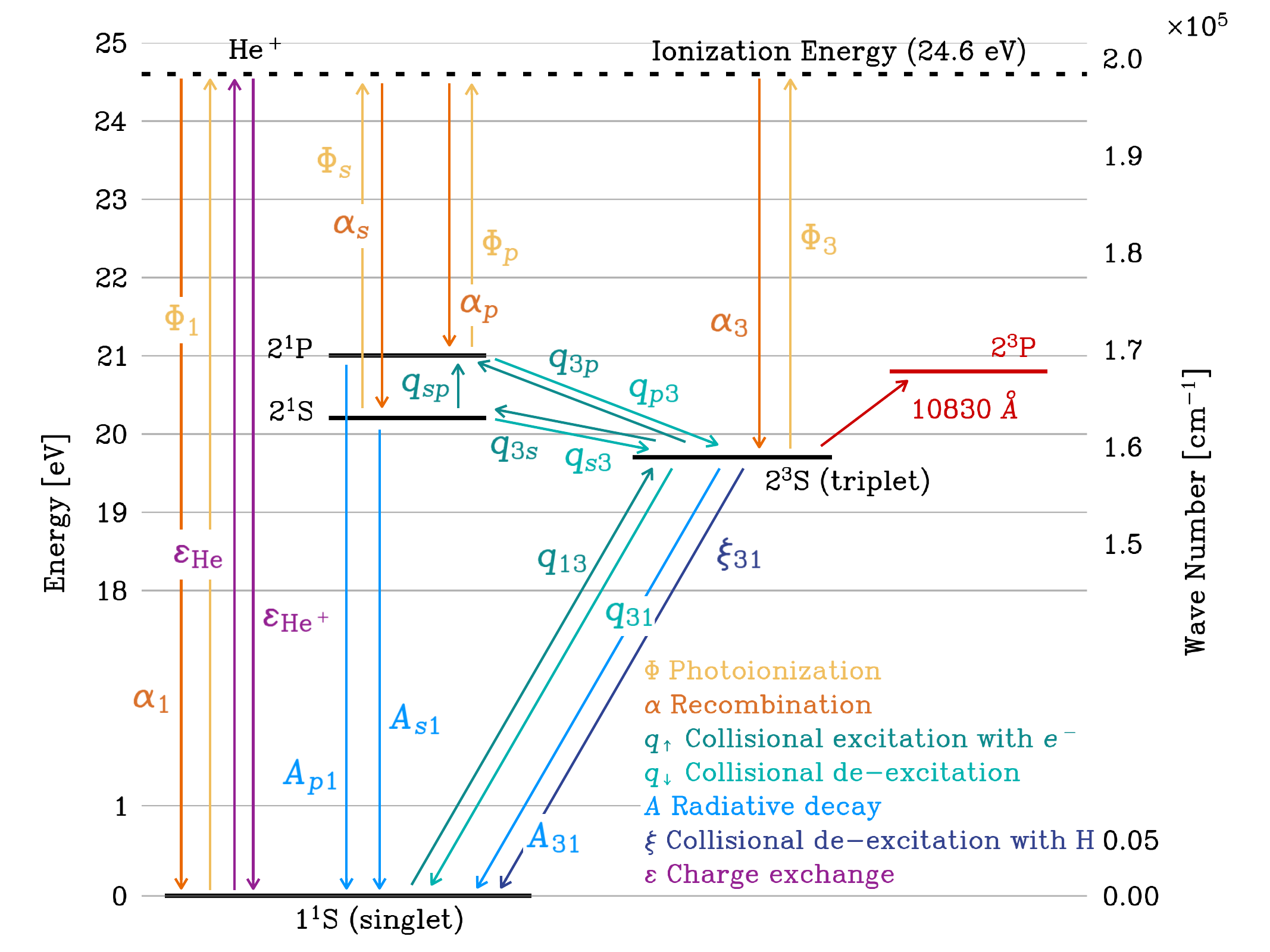}
    \caption{Atomic levels of helium, illustrating the transitions considered in our calculations, including transitions from \protect\cite{2024arXiv241205258S}. For completeness, we show all the transitions, even though we ignore those populating/depopulating the 2$^{1}$P state.}
    \label{fig:grotrian}
\end{figure*}

\subsection{Boundary Conditions} \label{sec:bound_cond}
The majority of the incoming high-energy photons are absorbed by the planet within a layer of gas at a specific radius, where the optical depth to XUV radiation reaches unity. We define $R_{\rm XUV}$ as the radius at which the gas becomes optically thick to the incoming X-rays and UV (XUV) radiation and adopt: 
\begin{equation}
    \tau = \int_{R_{\rm XUV}}^\infty \sigma_{\rm XUV} n_{\rm H}(r) dr = 1.
    \label{eq:optical_depth_photo}
\end{equation}
$n_{\rm H}$ is the number density of neutral hydrogen atoms and \revision{$\sigma_{\rm XUV} \sim 2 \times 10^{-18} {\rm cm}^{-2}$ is the hydrogen} absorption cross-section to XUV radiation (appropriate for $\sim 20$eV photons, \citealt{MC2009}). \revision{The atmosphere is initially set to be composed of 90\% of hydrogen and 10\% of helium. Thus, our models are applicable for hydrogen-dominated atmospheres, rather than high-metallicity atmospheres.}
The energy absorbed from the incoming radiation is converted into thermal energy ($P {\rm d}V$ work) to escape the gravitational potential of the planet. The energy balance gives an expression for the so-called ``energy-limited'' mass-loss rate \citep{Baraffe2004},
\begin{equation}
    \Dot{M}_{\rm EL} = \epsilon \frac{F_{\rm XUV} R_{\rm XUV}^2 R_{\rm p}}{G M_{\rm p}},
    \label{eq:mloss_el}
\end{equation}
where $\epsilon$ is the efficiency parameter \revision{and is taken as a constant $\epsilon = 0.1$}. 
Once we have computed the radial density and velocity profiles, we calculate the mass-loss rate
\begin{equation}
    \Dot{M} = 4 \pi r^2 \varv(r) \rho(r),
\end{equation}
and we set at $r = R_{\rm XUV}$
\begin{equation}
    4 \pi R_{\rm XUV}^2 \rho(R_{\rm XUV}) \varv(R_{\rm XUV}) = \epsilon \frac{F_{\rm XUV} R_{\rm XUV}^2 R_{\rm p}}{G M_{\rm p}}.
    \label{eq:mloss_el_rxuv}
\end{equation}
The gas will behave as a transonic wind when $R_{\rm XUV}$ sits inside the sonic radius $R_{\rm s}$. Otherwise, the flow rapidly becomes supersonic. In such cases, we set the velocity to be $c_s$ at $R_{\rm XUV}$ and use the generalised Parker wind solutions from \citet{Owen2024}. 

\noindent Finally, we impose momentum conservation between the hydrostatic layer and the outflow, 
\begin{equation}
    \rho_{\rm hs} \, c_{s, {\rm eq}}^2 = \rho_{\rm p} \, (c_s^2 + \varv_{\rm p}^2). 
    \label{eq:mom_cons_rxuv}
\end{equation}
For every planet, given its mass $M_{\rm p}$ and radius $R_{\rm p}$, we self-consistently solve the system of equations~\ref{eq:optical_depth_photo}, \ref{eq:mloss_el_rxuv} and \ref{eq:mom_cons_rxuv} and find the values of $R_{\rm XUV}$, $T_{\rm w}$ and $\Dot{M}$ for which the equations are satisfied. This root-finding problem is numerically solved by employing the \textsc{brentq} method, provided in \texttt{SciPy} \citep{2020SciPy-NMeth}. Furthermore, following \citet{Owen2024} if this procedure yields a wind temperature over $10^4$~K, at which Ly-$\alpha$ cooling would thermostat the outflow back down to $10^4$~K \citep[e.g.][]{MC2009,Owen2016}, we assume the outflow is in radiative recombination equilibrium and cap the outflow temperature to $10^4$~K, reducing the mass-loss efficiency in the process.

\subsection{Stellar Properties} \label{sec:star}

For a given spectral type, we select the corresponding stellar spectrum from the MUSCLES survey \citep{2016ApJ...820...89F, 2016ApJ...824..101Y, 2016ApJ...824..102L}. Each spectrum corresponds to a specific star with measured mass and radius. We integrate the whole spectrum to calculate the bolometric luminosity, which is then used to determine the stellar effective temperature using
\begin{equation}
    L_* = 4 \pi R_*^2 \sigma_{\rm SB} T_{\rm eff}^4,
\end{equation}
where $R_*$ is the stellar radius and $\sigma_{\rm SB}$ is the Stefan-Boltzmann constant. The planet's equilibrium temperature is given by
\begin{equation}
    T_{\rm eq} = (1-A_{\rm B})^{1/4} T_{\rm eff} \sqrt{\frac{R_*}{2 a}},
    \label{eq:temp_eq}
\end{equation}
where $T_{\rm eff}$ is the stellar effective temperature and $a$ is the planet's semi-major axis. $A_{\rm B}$ is the Bond albedo, which we set to $0.3$ for simplicity.

\subsection{Hydrogen Distribution} \label{sec:hydrogen}
Given that the atmosphere predominantly consists of hydrogen, we first calculate the radial density profile for neutral and ionized hydrogen. We assume that the total number fraction of hydrogen is 0.9 and the remaining is the number fraction of helium.
We solve the equation of recombination/ionization balance \citep[see,][]{2018ApJ...855L..11O}
\begin{equation}
    \varv \frac{\partial f_{\rm ion}}{\partial r} = (1 - f_{\rm ion}) \Phi e^{-\tau_0} - n_{\rm H}f_{\rm ion}^2 \alpha_{\rm rec},
    \label{eq:dfion_dr}
\end{equation}
where $f_{\rm ion} = n_{\rm H^+} / (n_{\rm H^0} + n_{\rm H^+})$ is the fraction of ionized hydrogen and $n_{\rm H}$ is the hydrogen number density. Following \cite{2018ApJ...855L..11O}, we define the photoionization rate as
\begin{equation}
    \Phi = \int_{\nu_0}^\infty \frac{F_{\nu}}{h\nu} a_{\nu} d\nu
\end{equation}
and the optical depth as
\begin{equation}
    \tau_0 = a_0 \int_r^\infty n_{\rm H^0} dr, 
\end{equation}
where $a_0$ is the flux-averaged hydrogen photoionization cross section. The hydrogen recombination rate at any given temperature is given by
\begin{equation}
    \alpha_{\rm rec} = 2.59 \times 10^{-13} \left(\frac{T}{10^4}\right)^{-0.7} {\rm cm}^3 {\rm s}^{-1}.
\end{equation}
We initially assume that the atmosphere is completely neutral and calculate the optical depth, and then solve the differential equation by employing a Python ODE solver (\verb|solve_ivp|, using either the \verb|LSODA| or \verb|Radau| method).

\subsection{Helium Metastable State Distribution} \label{sec:helium}
The properties of the helium triplet population strongly depend on the incident stellar flux \citep{2019ApJ...881..133O}. The incoming radiation is strong enough to ionize the hydrogen and helium atoms in the upper layer of the atmosphere. Hydrogen ionization occurs at 912~\AA, while the He singlet and triplet ionization occurs at wavelengths 504~\AA~and 2593~\AA, respectively. We integrate the stellar spectrum over the wavelength range of ionizing photons to calculate the corresponding fluxes. 

To solve for the fraction of helium in the singlet 1$^{1}$S and triplet 2$^{3}$S states, we \revision{include the transitions listed in \cite{2024arXiv241205258S} in our calculations. We also follow the approach of \cite{2018ApJ...855L..11O}, \cite{2024MNRAS.527.4657A}, where the solution for the fraction of helium in the 2$^{1}$P state is not treated explicitly, following the discussion by \citet{2024arXiv241205258S}.} 
\revision{Due to the rapid decay of the 2$^{1}$P state into the 1$^{1}$S state ($f_p A_{p1}$), we allow electron collisions to populate the singlet state, as also done in \cite{2024MNRAS.527.4657A}. For the same reason, recombination into the 2$^{1}$P state contributes to populating the 1$^{1}$S state ($f_{\rm He^+}n_e\alpha_p$).}

Considering that $f_{\rm He^+} = (1 - f_1 - f_3 - f_s)$, we obtain
\begin{equation}
\begin{split}
    \varv \frac{\partial f_1}{\partial r} = \,
    & f_{\rm He^+}n_e\alpha_1 - f_1\Phi_1e^{-\tau_1} + f_3 A_{31} + f_s A_{s1} \\
    & + f_3 n_e q_{31} - f_1 n_e q_{13} \\ 
    & + f_3 n_{\rm H^0}\xi_{31} + f_{\rm He^+}n_{\rm H^0}\varepsilon_{\rm He^+} - f_1n_{\rm H^0}\varepsilon_{\rm He} ,
\end{split}
\label{eq:df1_dr}
\end{equation}
\begin{equation}
\begin{split}
    \varv \frac{\partial f_3}{\partial r} = \,
    & f_{\rm He^+}n_e\alpha_3 - f_3 \Phi_3 e^{-\tau_3} - f_3 A_{31} + f_1 n_e q_{13} \\ 
    & + f_s n_e q_{s3} - f_3 n_e q_{3s} - f_3 n_e q_{3p} \\ 
    & - f_3 n_e q_{31} - f_3 n_{\rm H^0} \xi_{31},
\end{split}
\label{eq:df3_dr}
\end{equation}
\begin{equation}
\begin{split}
    \varv \frac{\partial f_s}{\partial r} = \,
    & f_{\rm He^+}n_e\alpha_s - f_s \Phi_s e^{-\tau_s} - f_s A_{s1} \\
    & + f_3 n_e q_{3s} - f_s n_e q_{s3} - f_s n_e q_{sp} .
\end{split}
\label{eq:dfs_dr}
\end{equation}
Each of the terms in Eqs.~\ref{eq:df1_dr}, \ref{eq:df3_dr} and \ref{eq:dfs_dr} are indicated by the arrows in Figure~\ref{fig:grotrian}. We assume that all the helium is initially in the ground state and that the number density of electrons is equal to the number density of hydrogen ions. Additionally, we include the contribution of neutral hydrogen to the total optical depth. 
Similarly to the calculation of the hydrogen fraction, we use a Python ODE solver (\verb|solve_ivp|, using either the \verb|LSODA| or \verb|Radau| method) to compute the fractions of helium in the triplet, singlet and excited $2^1$S states. 



\subsection{Transit} \label{sec:transit}
We simulate a transit to compute the absorption of the stellar spectrum due to the presence of the planetary atmosphere. 
First, we construct a 2D grid which extends along the line of sight $x$ and along the atmosphere's height in the vertical direction $r$. 
Then, we compute the de-projected density $\rho_{\rm los}$ and velocity $\varv_{\rm los}$ along the line of sight and calculate the overall optical depth. The contribution from the entire atmosphere is then determined by summing across the vertical extend out to the Coriolis radius, as described in Section~\ref{sec:overview}. 
We recall Eq.~\ref{eq:excess_absorption} to estimate the excess absorption from our model, where the optical depth is therefore defined as \citep[see, e.g.,][]{2010ApJ...723..116K, 2018ApJ...855L..11O}
\begin{equation}
    \tau_{\nu}(b) = 2 \int_b^{\infty} \frac{n_3(r) \sigma_0}{\sqrt{r^2 - b^2}} \Phi(\Delta \nu) r dr.
\end{equation}
Here, we have included the line-broadening term where we consider a Voigt profile $\Phi(\Delta \nu)$, where $\Delta \nu = \nu-\nu_0 + (\nu_0/c) \varv_{\rm los}$.
The corresponding half width at half maximum for the two curves is given by \citep[e.g.,][]{2020A&A...636A..13L, 2022A&A...659A..62D}
\begin{equation}
\begin{split}
    \alpha = & \frac{\nu_0}{c} \sqrt{\frac{2 k_B T}{m_{\rm He}}}, \\
    \gamma = & \frac{A_{ij}}{4 \pi},
\end{split}
\end{equation}
where $A_{ij}$ is the Einstein coefficient for the transition. The Voigt profile is computed using the corresponding Python module in the \verb|scipy| package. 


\section{Results and Discussion} \label{sec:results_discuss}
Our model is not as accurate as the latest simulations \citep[e.g.][]{Linssen2022,2024MNRAS.527.4657A, biassoni2024}; however, it is more physically consistent than approaches where the outflow temperature and mass-loss rate are unlinked \citep[e.g.][]{2018ApJ...855L..11O,dosSantos2022}. Thus, the advantage of our model is that it is simple enough to allow us to explore the parameters range extensively while still incorporating basic physics given the importance of self-consistently linking temperature to mass-loss rate (Section~\ref{sec:overview}). Once we have calculated the excess absorption, we derive the corresponding equivalent width of each spectrum. The equivalent width is then scaled by the \textit{geometric} factor following Eq.~\ref{eq:scaled_excess_abs}. 
Based on our calculations in Section~\ref{sec:overview}, the equivalent width is expected to scale linearly with the mass-loss rate, with a non-linear correction due to the temperature. 

\begin{figure}
    \centering
    \includegraphics[width=\columnwidth,trim=0cm 0cm 0cm 0cm,clip]{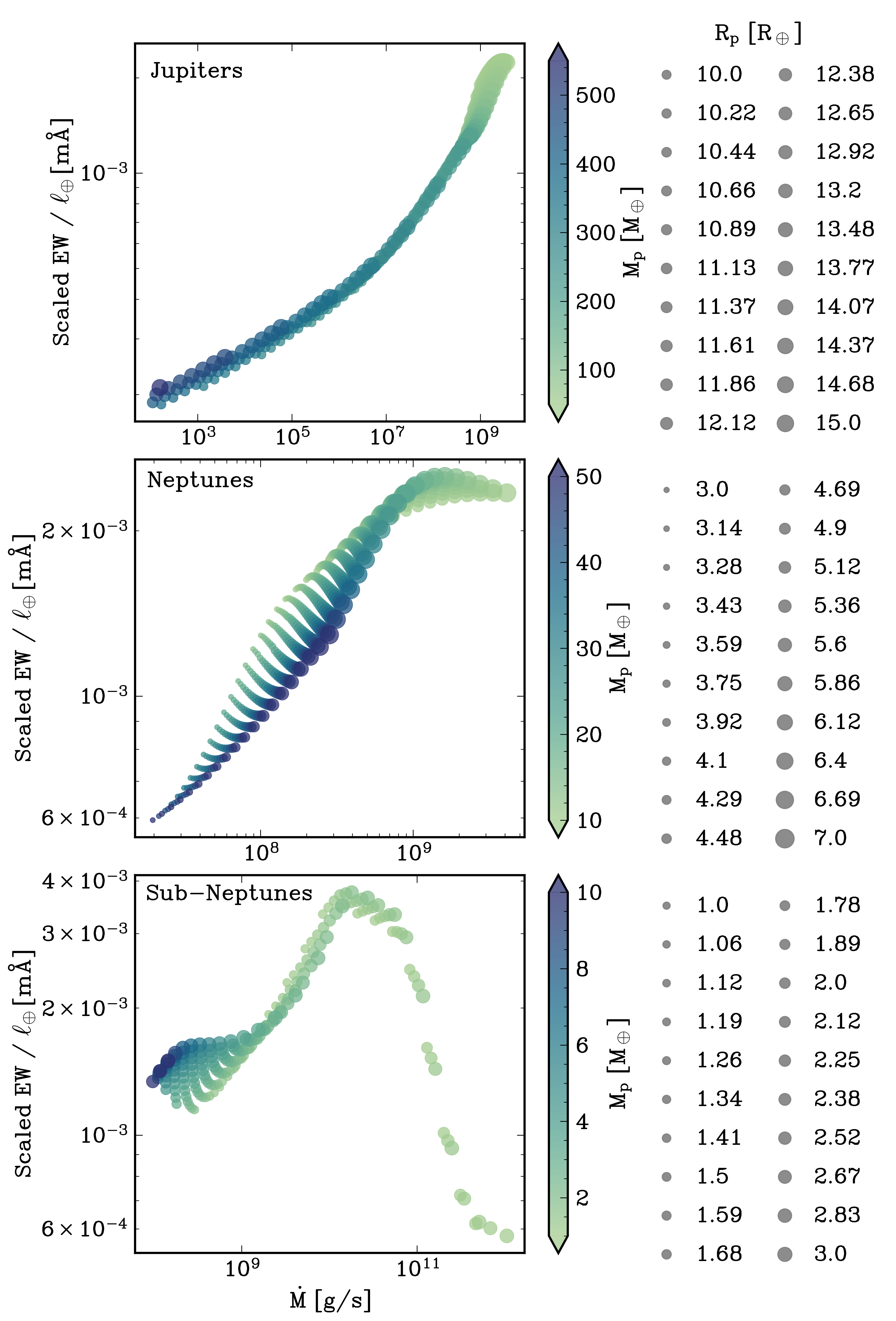}
    \caption{Scaled equivalent width as a function of mass-loss rate, for Jupiters, Neptunes and Sub-Neptunes. The points colour indicates the planet's mass and the marker size scales with the planet's radius, whose values are listed in the legend. Each exoplanet is located at 0.05~au and illuminated by an M2.5 type star.}
    \label{fig:results_mp}
\end{figure}
\begin{figure*}
    \centering
    \includegraphics[width=\textwidth,trim=0cm 0cm 0cm 0cm,clip]{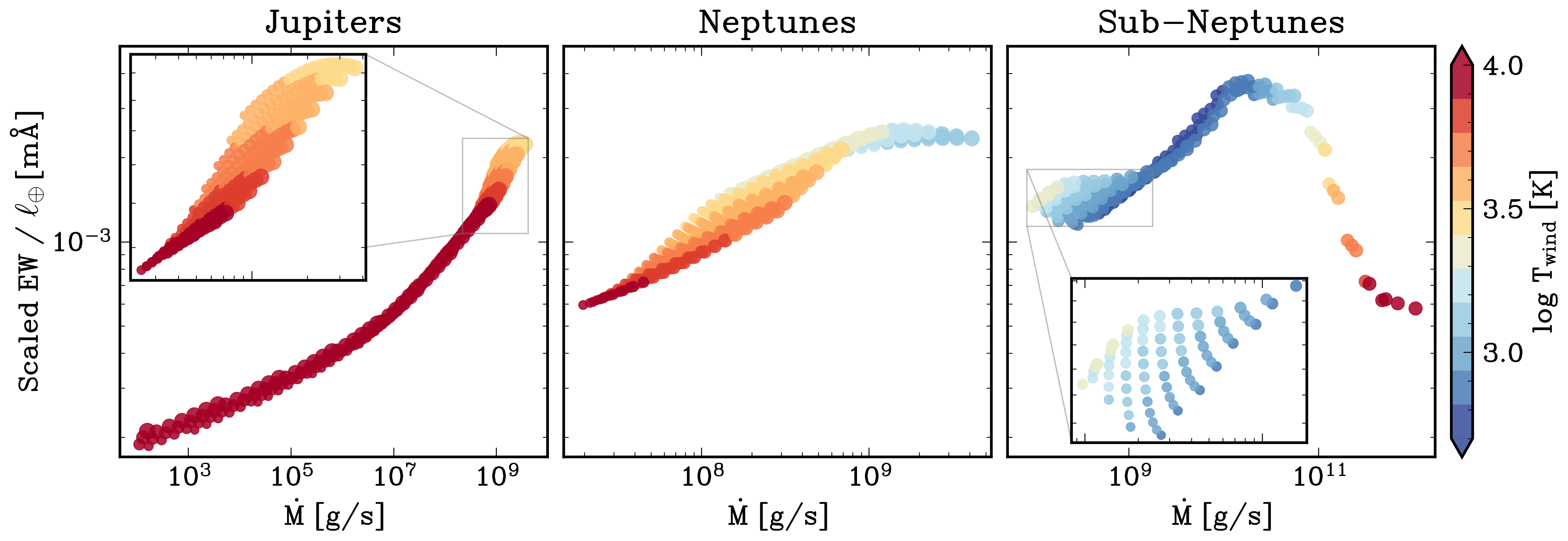}
    \caption{Scaled equivalent width as a function of mass-loss rate, colour-coded by the outflow temperature. The marker size scales with the planet radius, whose values are listed in the legend of Figure~\ref{fig:results_mp}. Each panel shows the results for the three different populations of exoplanets, Jupiters, Neptunes and Sub-Neptunes. The first and third panels show also a zoom-in onto the rectangular region to highlight the dependency from the wind temperature.}
    \label{fig:results_twind}
\end{figure*}
\begin{figure*}
    \centering
    \includegraphics[width=\textwidth,trim=0cm 0cm 0cm 0cm,clip]{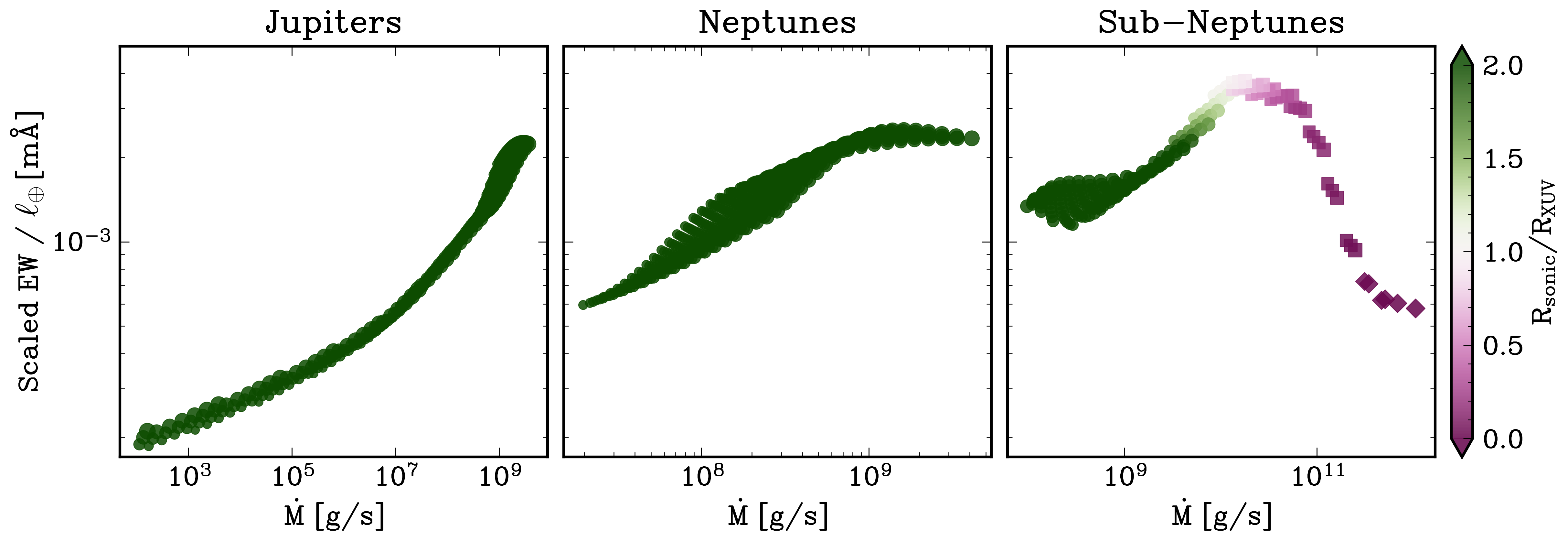}
    \caption{Scaled equivalent width as a function of mass-loss rate, for Jupiters, Neptunes and Sub-Neptunes. The points colour represents the ratio between the (\textit{hot}) sonic radius and $R_{\rm XUV}$ and the marker size scales with the planet radius, whose values are listed in the legend of Figure~\ref{fig:results_mp}. Squares indicate planets for which $R_{\rm XUV}$ is larger than the \textit{hot} sonic radius, while diamonds represent planets for which $R_{\rm XUV}$ is larger than the \textit{cold} sonic radius and are therefore in the core-powered mass-loss regime (see Section~\ref{sec:atmosphere_structure} for further details).}
    \label{fig:results_ratioradii}
\end{figure*}

\subsection{The general trend} \label{sec:trend}
As will become clear, different planet types sit in different regions of parameter space. Thus, we consider three populations of exoplanets: Sub-Neptunes, Neptune and Jupiter-like planets. We run our model over a grid of 400 exoplanets for each population, spanning over a range of 20 values in both masses and radii. The range of values considered for each population is listed in the top section of Table~\ref{tab:param_space}. In order to avoid obtaining a planet's properties that are unphysical, we constrain the planet's density to values lower than \revision{3} g~cm$^{-3}$. Each exoplanet is located at 0.05~au and irradiated by an M2.5 type star. 
We plot the results from our models in Figure~\ref{fig:results_mp}, which shows the scaled equivalent width as a function of mass-loss rate. Each row shows the results for the three different populations of exoplanets. The colour indicates the planet's mass, and the marker size scales with the planet's radius (whose values are listed in the legend for each panel). 
The general trend is similar amongst different populations, as it shows an overall correlation between helium absorption and increasing mass-loss rate, as expected from our discussion earlier. The details of the behaviour shown in each distribution are investigated below. 
\begin{table}
    \centering
    \caption{\textit{Top:} Range of planetary masses and radii considered for each population of exoplanets. \textit{Bottom:} Range of orbital separations and spectral types considered in this study.}
    \begin{tabular}{l|c|c}
    \toprule
        & \multicolumn{1}{c}{Mass [M$\earth$]} & \multicolumn{1}{c}{Radius [R$\earth$]} \\
    \midrule
        Sub-Neptunes & [1, 10] & [1, 3] \\
        Neptunes & [10, 50] & [3, 7] \\
        Jupiters & [50, 550] & [10, 15] \\
    \midrule
        \multicolumn{1}{l}{Orbital separation} & \multicolumn{2}{c}{0.01 - 0.1 au} \\
        \multicolumn{1}{l}{Spectral type} & \multicolumn{2}{c}{M8, M5.5, M2.5, K7, K2,} \\
         & \multicolumn{2}{c}{K1, G7, G2, G0, F8, F6} \\
    \bottomrule
    \end{tabular}
    \label{tab:param_space}
\end{table}

\begin{figure*}
    \centering
    \includegraphics[width=\textwidth,trim=0cm 0cm 0cm 0cm,clip]{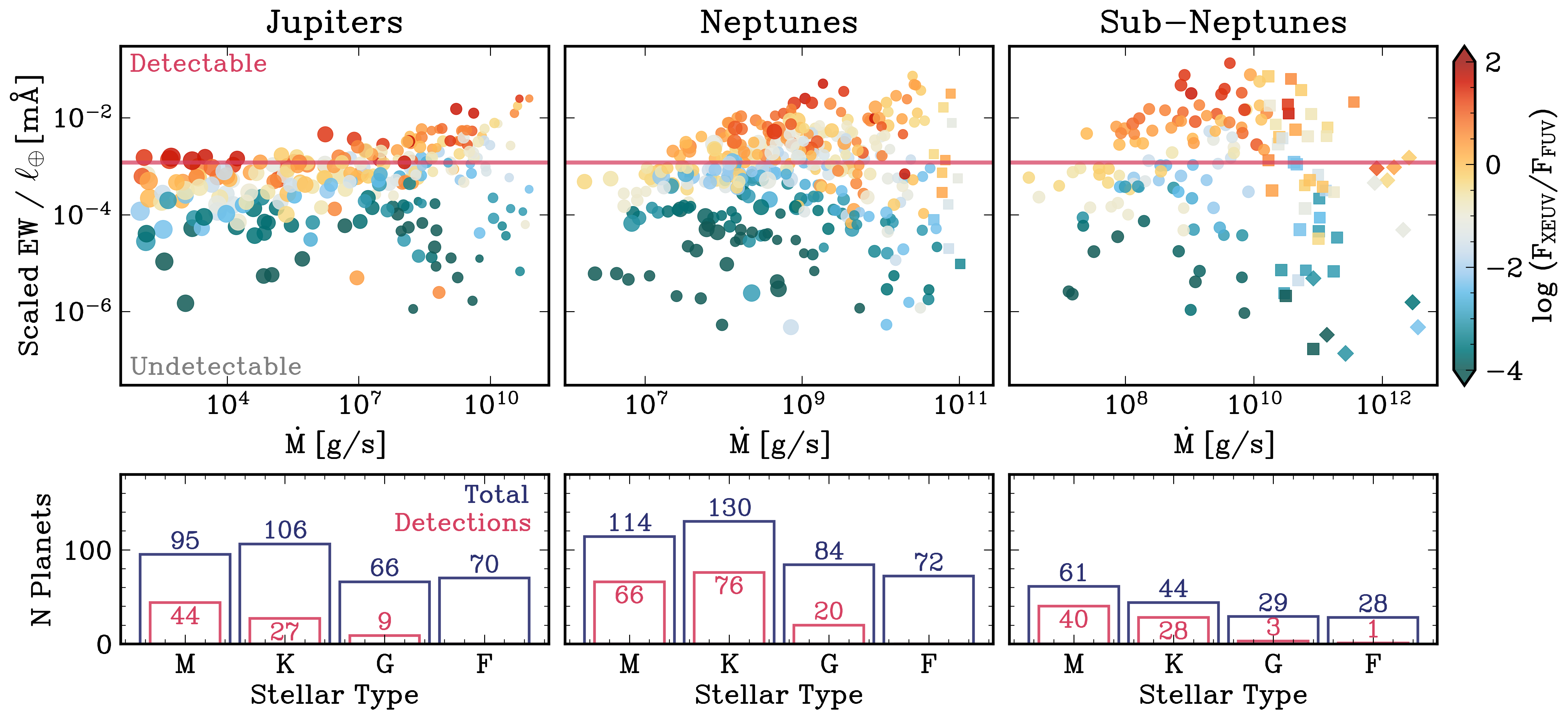}
    \caption{\textit{Top}: Scaled equivalent width as a function of mass-loss rate, for Jupiters, Neptunes and Sub-Neptunes. Each dot corresponds to an exoplanet with a mass and radius randomly selected amongst the values in Table~\ref{tab:param_space}. A random stellar spectra and orbital separation are assigned to each planet. 
    The colours scale with the ratio of XEUV to FUV fluxes and the marker size scales with the planet's gravity.  
    Following the convention introduced in Figure~\ref{fig:results_ratioradii}, squares indicate those planets for which $R_{\rm XUV}$ is larger than the \textit{hot} sonic radius, while diamonds represent planets for which $R_{\rm XUV}$ is larger than the \textit{cold} sonic radius (see Sections~\ref{sec:bound_cond} and \ref{sec:trend} for more details). 
    The red horizontal line indicates a typical detection threshold (see text for the details on the calculation). 
    \textit{Bottom}: Histogram of the frequency of exoplanets found around each stellar type (M, K, G and F). The blue columns indicate the total number of exoplanets, while the red columns show the number of detected exoplanets, meaning those ones above the detection threshold.}
    \label{fig:results_random_planets}
\end{figure*}

\paragraph*{Jupiters} have the highest gravitational pull on the gaseous atmosphere, and therefore, gas is only able to escape inefficiently. 
Jupiters reach lower values of mass-loss rates compared to the other population of exoplanets and lower helium absorption, as shown in the top panel of Figure~\ref{fig:results_mp}. For a fixed planetary mass, the excess absorption increases with increasing mass-loss rate and increasing planetary radius (which means lower gravity). The equivalent width also increases for lighter planets, which again means lower gravitational attraction of the gas in the atmosphere. 
The gas needs to be hot to be able to escape the gravitational potential of the Jupiters. We plot the temperature trend for the three populations of exoplanets in Figure~\ref{fig:results_twind}. The Jupiter-like planets are characterised by the highest wind temperatures (essentially, as they have the highest escape velocities), as shown in the left panel. \revision{This result has also already been found in previous studies, for example in \cite{2016A&A...586A..75S}.}
The inset shows the trend with the outflow temperature: the higher the temperature, the lower the helium absorption. While this might seem counter-intuitive as higher temperatures give more powerful outflow, it boils down to the balance between collisional de-excitations and recombinations, which are the dominant transitions that determine the abundance of helium in the metastable state (Section~\ref{sec:overview}, \citealt{2019ApJ...881..133O} \& \citealt{biassoni2024}). Given collisional depopulation is exponentially sensitive to temperature, with higher temperatures yielding faster depopulation, with increasing temperature, the helium triplet state is more and more depopulated by the collisional de-excitations. Our semi-analytic results match the predictions outlined in Section~\ref{sec:overview} and shown in Figure~\ref{fig:theoretical_behaviour}. 
The temperature in the wind reaches $10^4$~K for the most massive planets within the sample. This causes a flattening trend because the gas has reached the maximum amount of helium that can be collisionally de-excited from the metastable state.

\paragraph*{Neptunes,} with intermediate gravity, can lose their atmosphere more easily. The helium absorption increases with increasing mass-loss rate, for bigger and less massive planets. While the equivalent widths have values similar to those of the Jupiters, the mass-loss rates of the Neptunes are generally higher. This is expected from planets with gravity lower than giant planets. Again, we see that the absorption increases with increasing mass-loss rates for larger and lighter planets. 
The middle panel of Figure~\ref{fig:results_twind} shows the wind temperature for the Neptunes. The spread in the distribution varies with the temperature, following our theoretical predictions in Figure~\ref{fig:theoretical_behaviour}. Again, the trend shows increased helium absorption at cooler gas temperatures. 

\paragraph*{Sub-Neptunes} have a much lower gravity compared to the other exoplanets, and therefore, it is much easier for the gas to escape the gravitational potential. Consequently, the gas is generally cooler, as shown in the right panel of Figure~\ref{fig:results_twind}. The inset highlights again how the equivalent width increases for cooler gas temperatures. 
An interesting behaviour shows up for this population. For the higher-mass planets within the sample, the equivalent width is proportional to the mass-loss rate, and the trend generally follows the Neptune and Jupiter-like planets. However, there is an inversion in the trend for the Sub-Neptunes with the smallest gravity. To understand this behaviour, we plot the same distribution as a function of the ratio between the sonic radius and $R_{\rm XUV}$, shown in Figure~\ref{fig:results_ratioradii}. 
For such small exoplanets, the sonic radius is smaller than $R_{\rm XUV}$, as shown in the right panel. As anticipated in Section~\ref{sec:bound_cond}, those planets have rapidly expanded atmospheres. This rapid expansion, from sub-sonic to super-sonic velocities, results in high mass-loss rates and a strong density drop, which results in very diffuse outflows. These low densities then yield low helium absorption signals, despite the strong mass-loss rates, and could be the explanation for several of the non-detections found around very young, puffy planets \citep[e.g.][]{Alam2024}. \revision{We briefly underline here that some Sub-Neptunes likely have high-metallicity atmospheres with increased mean molecular weights ($\mu\sim10$, e.g., \citet{Benneke2024,Holmberg2024}) compared to the hydrogen-dominated ($\mu=2.3$) atmospheres assumed in our model. This higher $\mu$ leads to a more compact atmosphere with reduced helium column density, weakening the helium absorption signal. As a result, while our model provides general physical insights that are still valid, it may overestimate the detectability of helium absorption in high-metallicity sub-Neptunes. Future work should focus on incorporating more realistic atmospheric compositions.} \\

\noindent The main implication highlighted by these results is the presence of a correlation between the helium absorption features we detect in exoplanet atmospheres and properties of the escaping atmosphere, such as the mass-loss rate. The outflow temperature introduces a spread in the correlation, resulting in cooler outflows to produce larger equivalent widths. 
Therefore, it is possible to disentangle the mass-loss rate and the temperature of the escaping atmosphere with high-resolution measurements, where the width of the line can be used to constrain the temperature \citep[e.g.][]{dosSantos2022}. 
However, care should be taken when looking at specific exoplanets, as Jupiters and Sub-Neptunes are two distinct populations of exoplanets. They behave differently, mainly due to their properties, so they should be treated separately, as we will see in the next Sections.


\subsection{Variability with stellar properties}
\begin{figure}
    \centering
    \includegraphics[width=0.9\columnwidth,trim=0cm 0cm 0cm 0cm,clip]{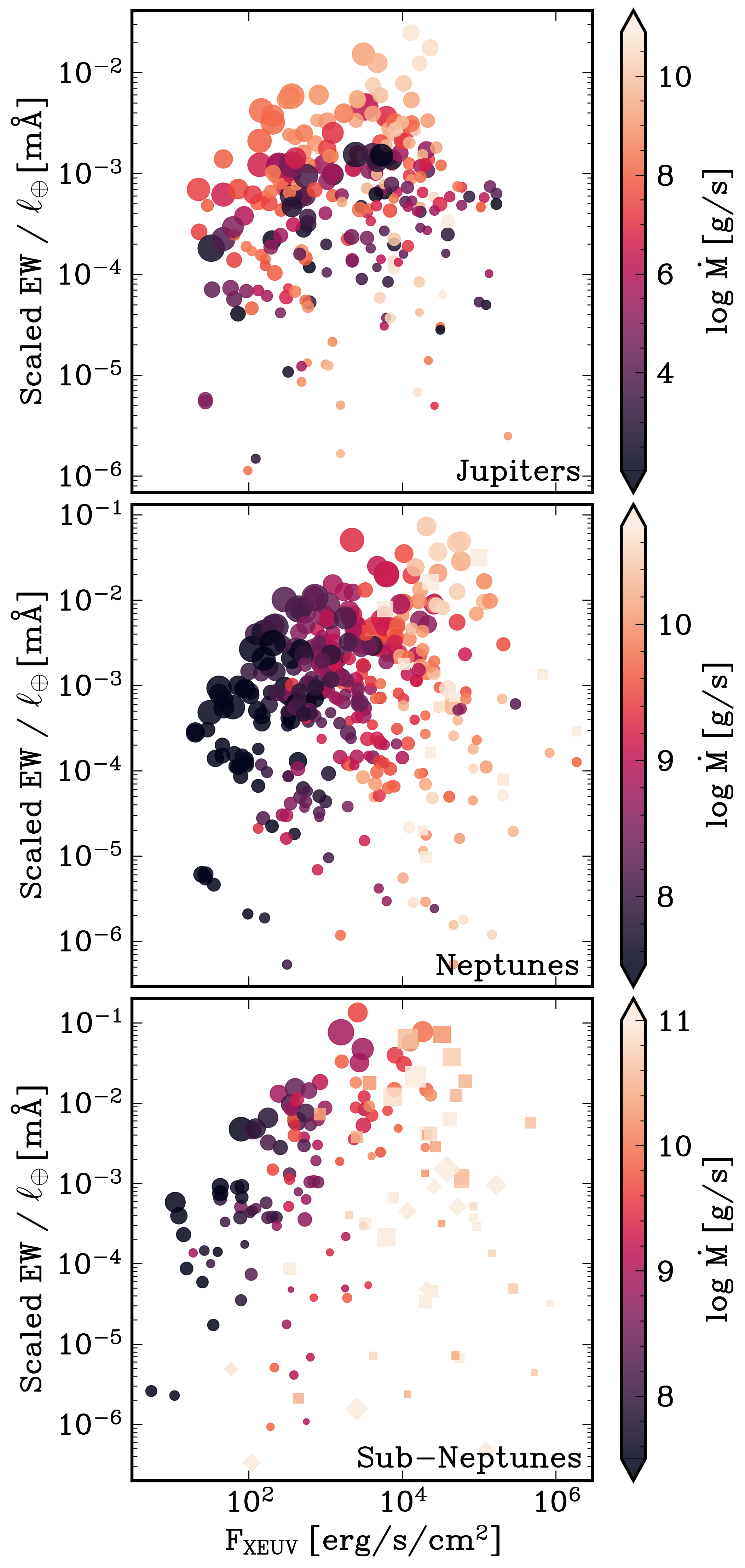}
    \caption{Distribution of XEUV fluxes and scaled equivalent widths, shown for 400 Jupiters, Neptunes and Sub-Neptunes (as indicated by the label in the top left corner). Each point is coloured by the mass-loss rate, and the maker size scales with the Coriolis radius. Different marker shapes are chosen following Figure~\ref{fig:results_ratioradii} (see also Sections~\ref{sec:bound_cond} and \ref{sec:trend} for more details).}
    \label{fig:results_xeuvflux}
\end{figure}
\begin{figure}
    \centering
    \includegraphics[width=0.9\columnwidth,trim=0cm 0cm 0cm 0cm,clip]{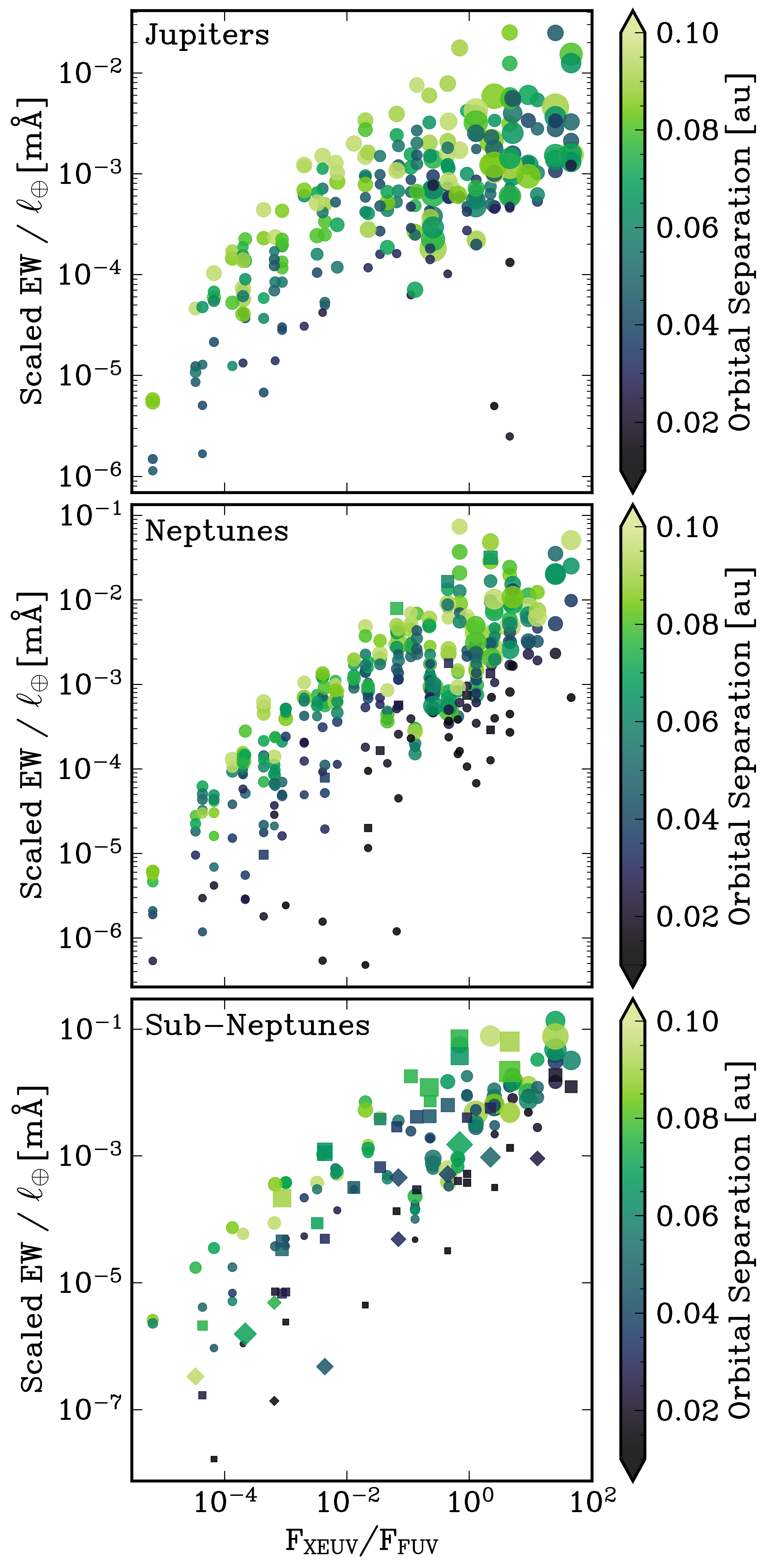}
    \caption{Correlation between the scaled equivalent width and the ratio between XEUV and FUV fluxes, shown for 400 Jupiters, Neptunes and Sub-Neptunes (as indicated by the label in the top left corner). Each point is coloured by the planet's orbital separation and the maker size scales with the Coriolis radius. Different marker shapes are chosen in accordance with Figure~\ref{fig:results_ratioradii} (see also Sections~\ref{sec:bound_cond} and \ref{sec:trend} for more details).}
    \label{fig:results_dist}
\end{figure}

The amount of helium found in the triplet state depends on the incoming stellar radiation. This has extensively been investigated in \cite{2019ApJ...881..133O} and \citet{biassoni2024}. They found that stellar spectra characterised by higher flux levels at small wavelengths ($\lambda$ < 1000~\AA) populate the helium triplet state more than stars with a lower XUV flux. On the contrary, stars with a higher flux in the mid-UV wavelength range will exhibit a lower degree of helium absorption, as they are less efficient at exciting the helium in the triplet metastable state. Harder spectra, therefore, induce stronger helium absorption, resulting in larger equivalent widths. 

\begin{figure*}
    \centering
    \includegraphics[width=\textwidth,trim=0cm 0cm 0cm 0cm,clip]{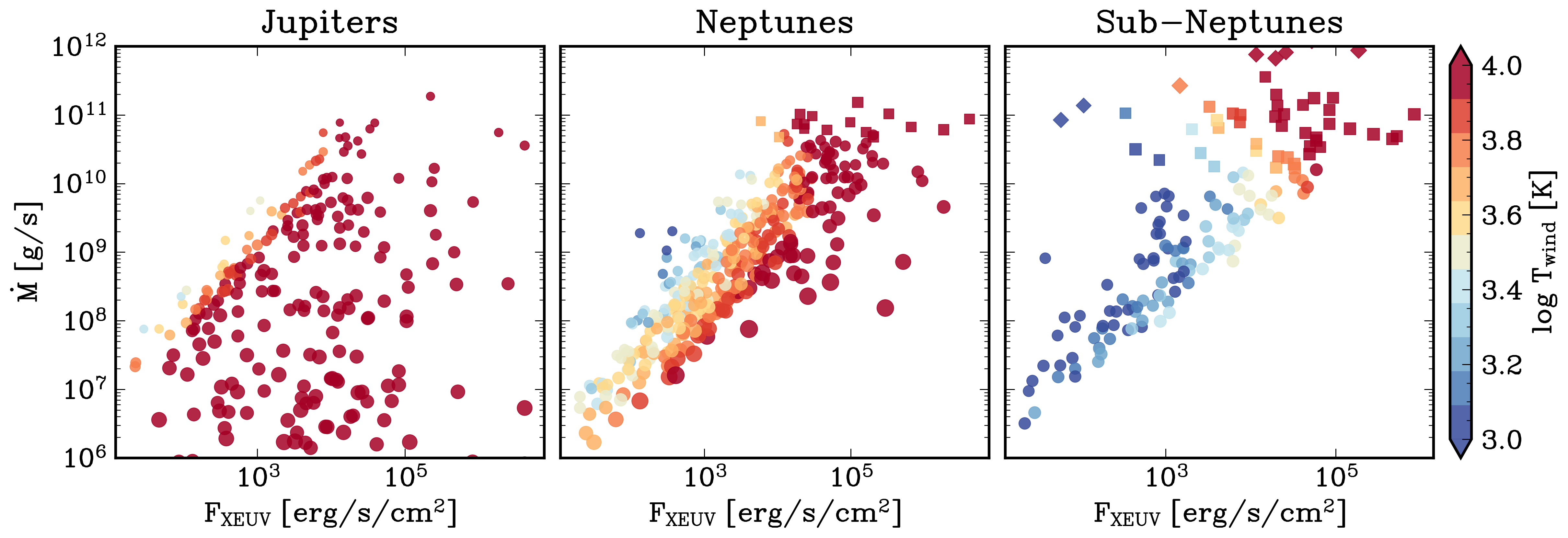}
    \caption{Correlation between XEUV fluxes and mass-loss rates, colour-coded based on the outflow temperature, shown for 400 Jupiters, Neptunes and Sub-Neptunes. The marker size scales with the planet's gravity. Different marker shapes are chosen in accordance with Figure~\ref{fig:results_ratioradii} (see also Sections~\ref{sec:bound_cond} and \ref{sec:trend} for more details).}
    \label{fig:results_mdot_flux}
\end{figure*}

We run a set of models to examine the effect of different stellar spectra. The models are run for 400 planets for each of the three populations, where the values of masses and radii are randomly drawn within the defined range for each population (see Table~\ref{tab:param_space} for details). Again, the planet's density is limited below \revision{3} g~cm$^{-3}$ to avoid unphysical values. 
A random stellar spectral type is also assigned to each planet and chosen from the following list: M8, M5.5, M2.5, K7, K2, K1, G7, G2, G0, F8, F6. A value of stellar mass and radius is then assigned based on the spectral type. The XEUV flux component of each stellar spectrum is also multiplied by a scaling factor randomly chosen amongst the values 0.1, 0.5, 1.0, 2.0, and 10.0 to maximise the spectral variability. 
Additionally, a random orbital separation between 0.01~au and 0.1~au (with steps of 0.005~au) is assigned to each planet. 

\begin{figure*}
    \centering
    \includegraphics[width=\textwidth,trim=0cm 0cm 0cm 0cm,clip]{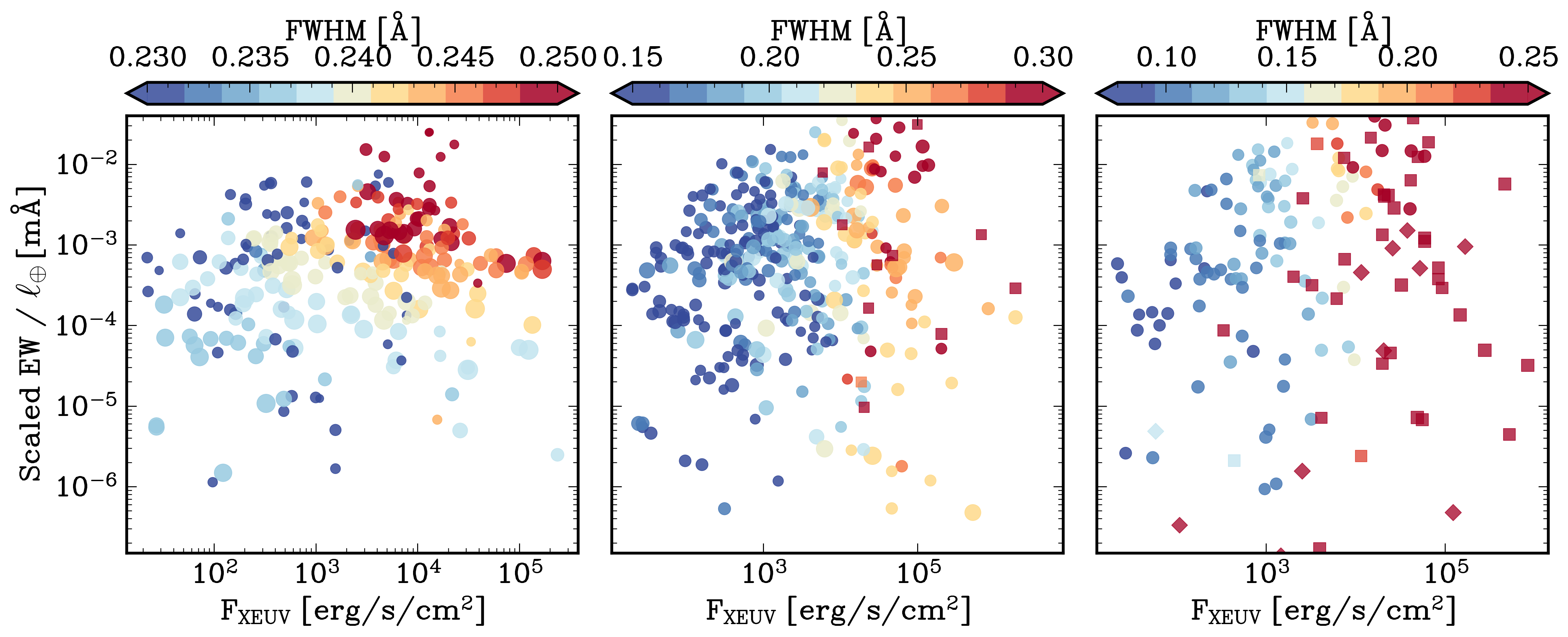}
    \caption{Distribution of scaled equivalent widths with XEUV flux, colour-coded based on the full width at half maximum. Each panel shows the results for 400 Jupiters, Neptunes and Sub-Neptunes, respectively. The marker size scales with the planet's gravity. Different marker shapes are chosen in accordance with Figure~\ref{fig:results_ratioradii} (see also Sections~\ref{sec:bound_cond} and \ref{sec:trend} for more details).}
    \label{fig:results_ew_fwhm}
\end{figure*}

The results are plotted in Figure~\ref{fig:results_random_planets}. The top row shows the scaled equivalent width as a function of mass-loss rate for Jupiters, Neptunes and Sub-Neptunes, with a significant scatter introduced by the stellar variability. We do not find a clear correlation between the mass-loss rate and the equivalent width. \revision{This result also aligns with the recent findings of \cite{2024arXiv241003228L}, which are based on a model incorporating a more detailed radiative transfer treatment. Their study similarly identifies only a weak correlation between the strength of the helium signal and the XUV flux, a result also found in this work (see Section~\ref{sec:xeuv_correlation} and Figure~\ref{fig:results_xeuvflux}).}
The markers are colour-coded based on the ratio of XEUV to FUV flux and the size scales with the planet's gravity. As expected, stars with harder spectra produce a higher absorption signal (and therefore larger equivalent widths), while planets with smaller gravitational attraction will induce stronger mass-loss. 

The red horizontal line indicates a typical detection limit for helium absorption. We consider the spectral resolution of Keck, R=25,000, and a FWHM of 4 pixels per resolution element. Assuming a typical noise per pixel of $\sigma_{\rm noise} = 0.1$\%, the equivalent width associated with the minimum detectable signal is
\begin{equation}
    EW \approx 3 \sigma_{\rm noise} \Delta \lambda \times {\rm FWHM}.
\end{equation}
We obtain an $EW \approx 5$~m\AA. We then scale it considering an M2.5 type star (0.42 $M_{\odot}$, 0.43 $R_{\odot}$) and an orbital separation of 0.05~au, according to Eq.~\ref{eq:scaled_excess_abs}, and obtain a scaled equivalent width of $\approx 0.0012$~m\AA. 


The bottom row of Figure~\ref{fig:results_random_planets} shows the frequency of each population of exoplanets found around each spectral type, grouped in M, K, G and F (blue histograms). The red histograms highlight the frequency of detected exoplanets only above the estimated threshold. 
We note that the distribution peaks around M and K-type stars for all populations of exoplanets, while the signal produced by G and F-type stars is completely suppressed. 
\revision{The Sub-Neptunes have fewer detections due to the lower overall number of planets in the sample in the first place (which is limited by the density threshold).
The Neptunes have a higher number of detections compared to the Jupiter-like planets, which are harder to detect due to the generally lower absorption.}




\subsection{Correlations with XEUV flux} \label{sec:xeuv_correlation}

We expect the helium absorption to scale with the incident stellar flux. 
The results presented in \cite{2022MNRAS.512.1751P} highlight that helium ionization is driven specifically by the narrow-band EUV photons. In this work, in fact, they find that the helium absorption signal seems to directly correlate with the EUV flux, despite the statistically low number of sources. 

When looking at our results for a population of 400 exoplanets we find a weak correlation with both XUV and EUV fluxes, similarly to previous findings \citep[e.g., ][]{2018Sci...362.1388N, 2019A&A...629A.110A, 2022MNRAS.512.1751P}. We plot the helium absorption obtained from our models as a function of XEUV flux in Figure~\ref{fig:results_xeuvflux}: the points colour indicates the mass-loss rate experienced by the planet, which tends to be higher for larger equivalent widths, especially for the Jupiters. 
We note, in particular, that XEUV fluxes strongly determine the rate of mass lost by the atmosphere. This is particularly evident for Neptunes and Sub-Neptunes, which show a clear correlation between XEUV flux and mass-loss rate. The helium absorption signal from Jupiter-like planets is instead completely dominated by the extent of the outflow, as larger Coriolis turning radii produce a larger absorption. \\

Most notably, we find a much stronger correlation between the helium absorption and the ratio between the XEUV and the FUV components of the stellar flux. This is perhaps not surprising considering that it is the hardness of the spectra that is responsible for populating the helium metastable state \citep{2019ApJ...881..133O}. In fact, both the XEUV and the FUV parts of the spectrum control the number of excited helium atoms: the population of helium in the triplet state is maximized when the XEUV flux is maximized, and the FUV flux is minimized. 
Figure~\ref{fig:results_dist} illustrates the scaled equivalent width as a function of F$_{\rm XEUV}/$F$_{\rm FUV}$, for the three distinct populations of exoplanets. The helium excess absorption measured by the equivalent width increases with increasing flux ratio for all populations. The points are colour-coded by the planet's orbital separation. It might seem counter-intuitive that larger orbital separations can lead to larger absorption signals, but this can be understood when looking at the size of the outflow. By scaling the marker size by the Coriolis radius, in fact, we notice that planets with small Coriolis radii produce smaller absorption signals, despite being closer to the star (a result also seen in Lyman-$\alpha$ transits, \citealt{Owen2023,SchreyerLy2024}). This is because the Coriolis radius is proportional to the orbital separation $R_{\rm c} \propto a^{3/2}$, which sets the limit on the transit signal. \\

The stellar flux determines the wind temperature, and, in particular, the XEUV flux fixes the mass-loss rate. In Figure~\ref{fig:results_mdot_flux}, we highlight the dependency between the XEUV flux, the mass-loss rate and the outflow temperature. The marker size scales with the planet's gravity, with larger circles indicating a stronger gravitational force. The mass-loss rate directly correlates with the XEUV flux, as we have already hinted at from Figure~\ref{fig:results_xeuvflux}, and the spread of such correlation depends on the wind temperature.

\subsection{Unambiguously determining atmosphere properties}
A significant variability in the spectrum, therefore, introduces a variability in the outflow properties. As a consequence, the spread introduced by the stellar variability prevents us from finding any tight correlation between the helium absorption signal and the properties of the planet. 
Nonetheless, with high resolution observations it is possible to measure the width of the line precisely enough to derive the atmosphere properties. The velocity of the wind along the line of sight is responsible for the broadening of the absorption line. The full width at half maximum of the absorption line, hence, serves as a proxy for the temperature of the outflow. 
\revision{While this approach works well in spherically symmetric (1D) models, 3D simulations of individual planets with strong helium signals indicate that the complex kinematics of winds could influence the helium line profile. For example, \cite{2021ApJ...914...99W} and \cite{2024arXiv241019381N} demonstrate how the three-dimensional nature of winds - absent in 1D models - may alter the line profile and would make it more difficult to infer the temperature accurately from the line width. However, 3D simulations are computationally demanding, and our simplified approach is well-suited for achieving a deeper physical understanding of the problem at the population level.}

We fit a Gaussian function to the synthetic transmission spectra and derive the full width at half maximum (FWHM) of the helium absorption line for each planet in our populations. Figure~\ref{fig:results_ew_fwhm} illustrates the relationship between the scaled equivalent width and the XEUV flux, colour-coded by the FWHM. The FWHM increases with increasing XUV flux, indicating that the stellar flux sets the temperature of the outflow. 
Therefore, from high-resolution measurements, we can derive an estimate of both the line width (FWHM), which is a proxy for the wind temperature and the line equivalent width, deriving an estimate for the XUV flux. Using the correlation found in Section~\ref{sec:xeuv_correlation} and illustrated in Figure~\ref{fig:results_mdot_flux}, we can then hopefully determine the mass-loss rate from the XUV flux and the wind temperature. 

\subsection{Limitations}
In this work, we have constructed a model for an escaping atmosphere to directly observationally constrain atmospheric escape and its impact on shaping the exoplanet population. 
Our model is relatively simple; however, the simplicity was a deliberate choice to allow us to understand the main physics at play here. Nonetheless, we highlight some of the assumptions we have made:
\begin{itemize}
    \item We have assumed an energy limited dominated outflow, which ultimately sets the mass-loss rate in the model. In principle, in the future, this could be replaced by numerically determined efficiency factors.
    \item The outflow is described by an isothermal Parker wind solution. Although more complex and complete models of an escaping atmosphere exist, we consider a simpler approach and assume a constant temperature throughout the outflow. This approach has allowed us to link wind temperature to mass-loss rate self-consistently for the first time in a simple modelling framework. 
\end{itemize}

\section{Conclusions}
We have developed a 1D model for an escaping hydrogen-helium dominated atmosphere and looked at how observational properties scale with atmospheric characteristics. In particular, we have derived a scaled equivalent width of the helium absorption line using a \textit{geometric} factor that incorporates both planetary and stellar properties. Our main results are as follows. 
\begin{enumerate}
    \item We identified a positive correlation between the scaled equivalent width and the envelope mass-loss rate. Helium absorption measurements therefore serve as a valuable proxy for determining the mass-loss rate of an escaping atmosphere. Such a correlation also depends on the temperature of the outflow: specifically, a cooler outflow produces larger equivalent widths. 
    \item The correlation between mass-loss rate and equivalent width, however, shows significant scatter when taking into account the stellar spectrum variability. The radiation from the star, in fact, plays a role in determining the wind temperature and in fixing the energy limited mass-loss rate. 
    \item The scaled equivalent width correlates with the XEUV flux. However, it correlates even more strongly with the ratio of XUV to FUV flux. In fact, the relative flux intensity at XEUV wavelengths is more efficient at \revision{populating the helium metastable state} than the mid-UV and FUV components of the spectrum\revision{, which de-populate the triplet state by photoionization}. 
    \item The stellar flux determines the atmosphere properties and, in particular, the XUV flux sets the mass-loss rate and the wind temperature. 
\end{enumerate}
To conclude, high-resolution observations \revision{
have the potential to provide valuable insights into the mass-loss rate, the temperature of the outflow, and the XUV stellar flux through measurements of the line equivalent width and the line full width at half maximum. While these estimates are subject to uncertainties due to the simplifying assumptions inherent in the model, the results presented here are still important for constraining atmosphere properties from helium absorption, especially when applied at the population level.}
Additionally, independent measurements of Ly-$\alpha$ transits would further constrain the mass-loss rate, by probing the gas sound speed, itself a measure of the gas temperature \citep{SchreyerLy2024}.

\section*{Acknowledgements}

We thank the anonymous reviewer for comments and questions that improved the manuscript. This work was supported by the European Research Council (ERC) under the European Union’s Horizon 2020 research and innovation programme (Grant agreement No. 853022, PEVAP). JEO is supported by a Royal Society University Research Fellowship. This work benefited from the 2023 Exoplanet Summer Program in the Other Worlds Laboratory (OWL) at the University of California, Santa Cruz, a program funded by the Heising-Simons Foundation. For the purpose of open access, the authors have applied a Creative Commons Attribution (CC-BY) licence to any Author Accepted Manuscript version arising. 

\section*{Data Availability}

The data and software underlying this article will be shared on reasonable request to the corresponding author.



\bibliographystyle{mnras}
\bibliography{references} 




\appendix

\begin{figure}
    \centering
    \includegraphics[width=\columnwidth,trim=0cm 0cm 0cm 0cm,clip]{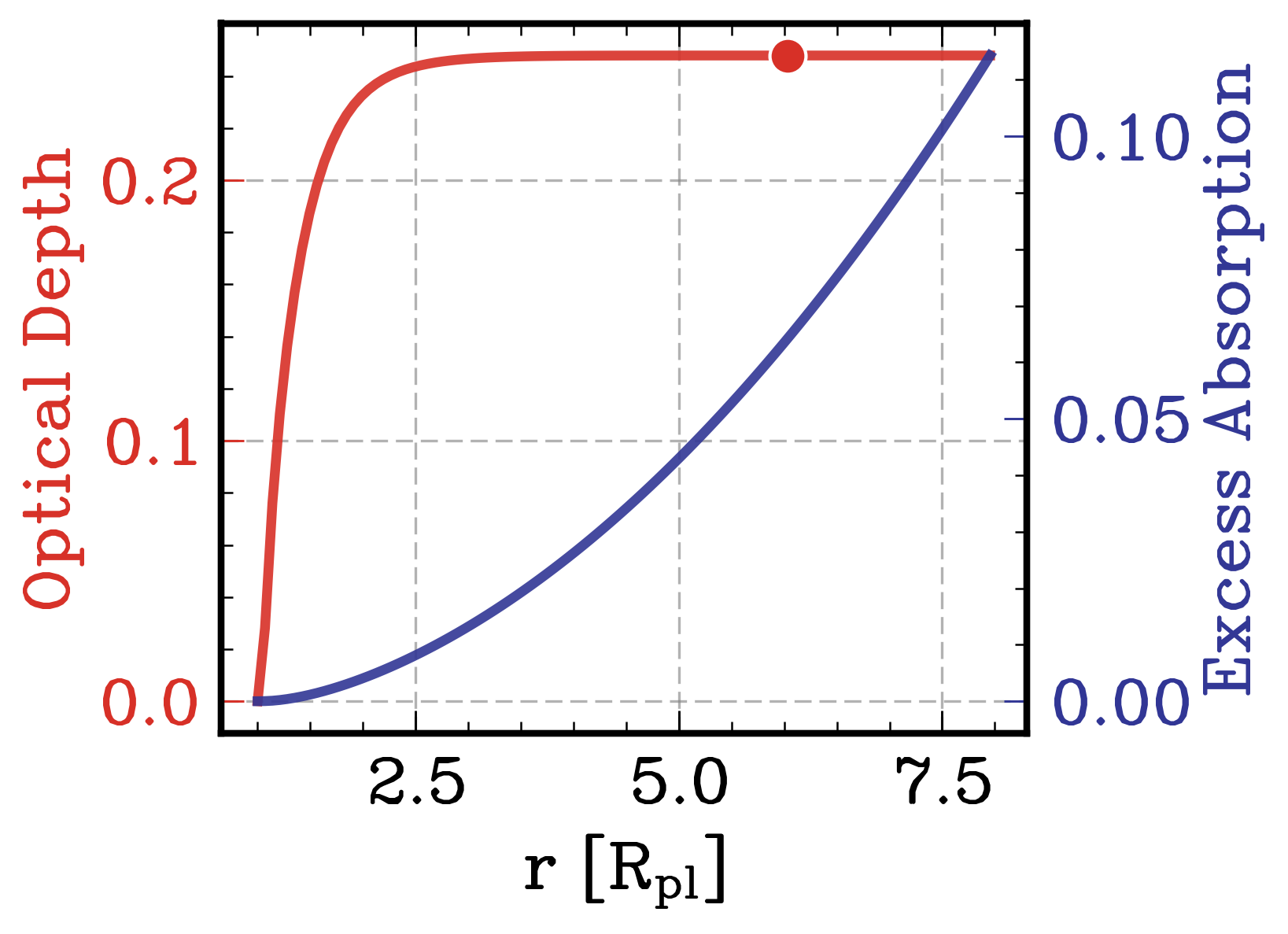}
     \caption{Optical depth in absorption (red curve), calculated as a cumulative integral along a ray from the base of the atmosphere to the Coriolis radius. The red dot indicates the sonic point. We also calculate the cumulative excess absorption (blue curve) in the atmosphere. This has been calculated using the model presented in Figure~\ref{fig:profiles}.}
    \label{fig:optical_depth}
\end{figure}

\revision{
\section{helium absorption is optically thin}
\label{appx:optical_depth_absorption}
Helium absorption is generally found to be optically thin \citep[e.g.][]{2023AJ....165...62Z, 2023ApJ...953L..25Z}. To illustrate this point, we use the definition of optical depth for helium absorption provided in Eq.~\ref{eq:optical_depth_abs}. The cumulative optical depth is calculated by integrating from the base of the atmosphere to the Coriolis radius. As shown in Figure~\ref{fig:optical_depth}, helium absorption remains predominantly optically thin. This conclusion holds for most of our models, supporting the assumptions made in this work. However, in a few cases, we find planets where the optical depth slightly exceeds unity, resulting in marginally optically thick absorption. Using this optical depth estimate, we compute the excess absorption, represented by the blue line in Figure~\ref{fig:optical_depth}. The results shown here also demonstrate that the excess absorption is primarily dominated at large radii, far from the planet's surface.}

\begin{figure}
    \centering
    \includegraphics[width=\columnwidth,trim=0cm 0cm 0cm 0cm,clip]{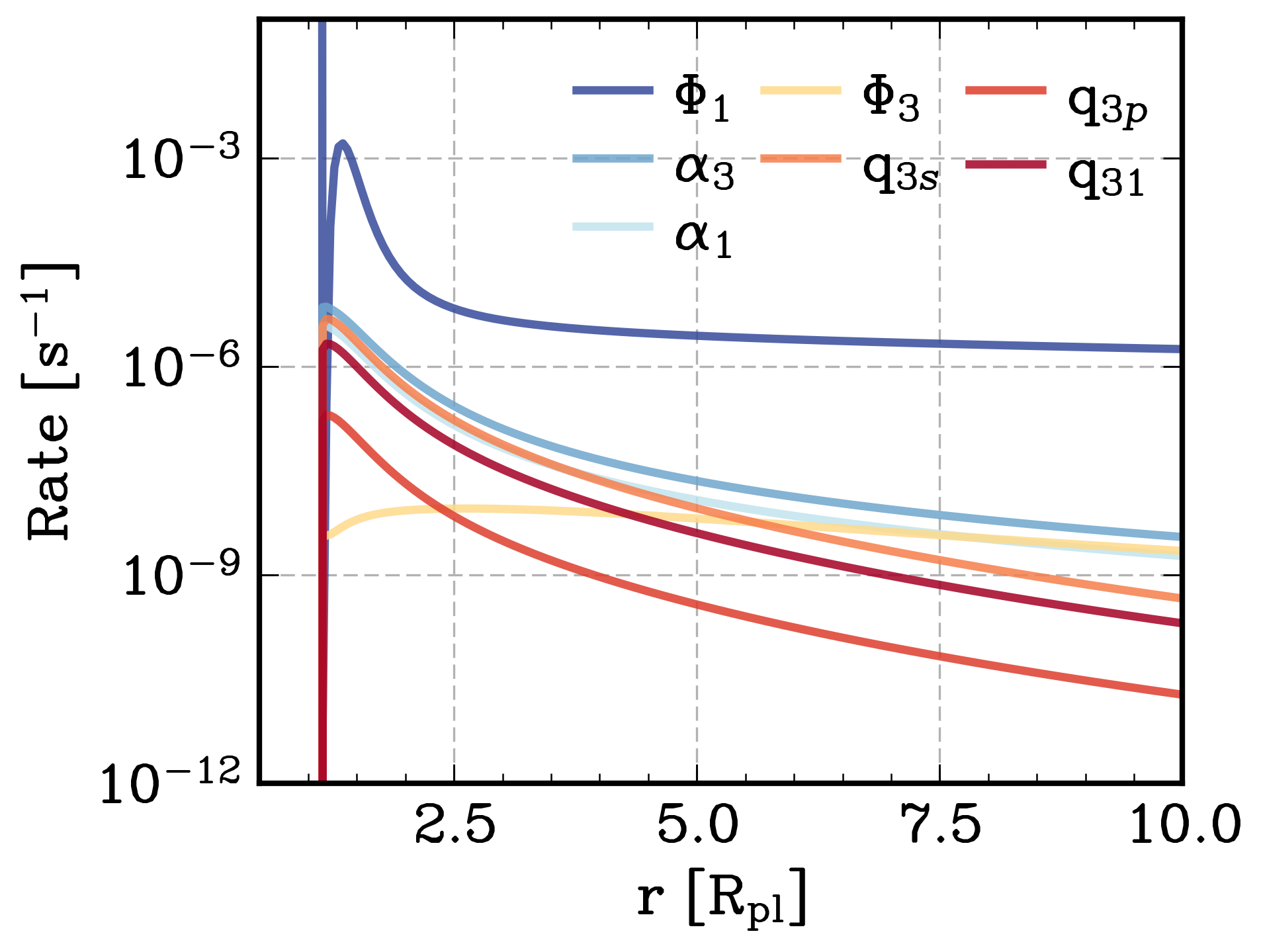}
     \caption{Photoionisation, recombination and collisional rates responsible for populating the helium triplet state. As in Figure~\ref{fig:optical_depth}, we use the model from Figure~\ref{fig:profiles} to produce these reaction rates. 
     Photoionization of helium produces a large fraction of ionized helium, which can then recombine equally in the singlet ($\alpha_1$) or triplet state ($\alpha_3$). Additionally, collisions with electrons can also increase the fraction of helium in the ground state (q$_{31}$), while decreasing the fraction of helium in the triplet state.}
    \label{fig:reaction_rates}
\end{figure}

\revision{
\section{the main reactions regulating the population of helium in the triplet state}
\label{appx:main_reactions}
Following the previous work by \cite{2019ApJ...881..133O}, we plot the magnitude of the main reaction rates responsible for determining the fraction of helium triplet in Figure~\ref{fig:reaction_rates}. The dominant reaction is photoionisation of the helium ground state ($\Phi_1$). The helium triplet is then produced by balancing recombination ($\alpha_3$), which populate the triplet, and collisional de-excitations with electrons (q$_{3s}$, q$_{3p}$ and q$_{31}$), which will depopulate the triplet to the singlet state. Photoionisation of the triplet state ($\Phi_3$) becomes important at larger distances (see \citet{2024arXiv241205258S}) and for stellar spectra with a strong FUV component – which is normally the case of G or Solar type stars. In this scenario, recombination into the triplet is balanced by photoionisation of the $2^3$S state. However, the resulting helium absorption is often too weak to produce a detectable signal \citep{2019ApJ...881..133O}.
}

\bsp	
\label{lastpage}
\end{document}